\documentclass[aps,prl,preprint,groupedaddress]{revtex4-2}

\usepackage{graphicx}
\graphicspath{{./figures/}}
\usepackage{newtxtext}
\usepackage{newtxmath}
\usepackage{natbib}
\usepackage[dvipsnames]{xcolor}
\usepackage{wasysym}
\usepackage{hyperref}
\usepackage{booktabs}
\usepackage{makecell}
\setcellgapes{3pt}        
\usepackage[left]{lineno}
\usepackage{array}
\renewcommand{\arraystretch}{0.6}
\renewcommand{\arraystretch}{0.6}            
\usepackage{array}
\newcolumntype{P}[1]{>{\raggedright\arraybackslash\setlength{\parskip}{0pt}}p{#1}}
\hypersetup{
    colorlinks = false,
    urlcolor   = black,
    citecolor  = black,
}

\newcommand{\RomanNumeralCaps}[1]
\linenumbers
\usepackage{tikz,xcolor,hyperref}
\usepackage{enumitem}
\definecolor{lime}{HTML}{A6CE39}
\DeclareRobustCommand{\orcidicon}{
	\begin{tikzpicture}
	\draw[lime, fill=lime] (0,0) 
	circle [radius=0.16] 
	node[white] {{\fontfamily{qag}\selectfont \tiny ID}};
	\draw[white, fill=white] (-0.0625,0.095) 
	circle [radius=0.007];
	\end{tikzpicture}
	\hspace{-2mm}
}
	
\foreach \x in {A, ..., Z}{\expandafter\xdef\csname orcid\x\endcsname{\noexpand\href{https://orcid.org/\csname orcidauthor\x\endcsname}
			{\noexpand\orcidicon}}
}

\newlist{inlineroman}{enumerate*}{1}
\setlist[inlineroman]{itemjoin*={{, and }},afterlabel=~,label=\roman*}

\draft 

\newcommand{\sixpointstar}{%
  \tikz[baseline=-0.1em,scale=0.2]{
    \draw[line width=0.5pt] (0,0.6) -- (-0.52,-0.3) -- (0.52,-0.3) -- cycle;
    \draw[line width=0.5pt] (0,-0.6) -- (-0.52,0.3) -- (0.52,0.3) -- cycle;
  }%
}

\begin{document}

\title{Filament inclination effect on turbulent canopy flows} 

\author{Shane Nicholas\orcidA{}}
\affiliation{Department of Engineering, City-St George's, University of London, Northampton Square, London EC1V 0HB, UK}
\author{Alessandro Monti\orcidB{}}
\author{Giulio Foggi Rota\orcidC{}}
\author{Marco E. Rosti\orcidD{}}
\affiliation{Complex Fluids and Flows Unit, Okinawa Institute of Science and Technology Graduate University (OIST), 1919-1 Tancha, Onna-son, Okinawa 904-0495, Japan}
\author{Mohammad Omidyeganeh\orcidE{}}
\author{Alfredo Pinelli\orcidF{}}
\affiliation{Department of Engineering, City-St George's, University of London, Northampton Square, London EC1V 0HB, UK}

\date{\today}

\begin{abstract}
Inspired by the spontaneous behaviour observed in filamentous layers—where the balance between flow-induced drag and structural elasticity dictates the filaments' equilibrium streamlined posture—we perform a series of large eddy simulations to investigate how filament inclination affects turbulent shear flows developing both above and within a canopy of filaments.

We examine six distinct filament inclination angles ranging from $0^{\circ}$ to $90^{\circ}$. The in-plane solid fraction and filament length are chosen to achieve a fully dense canopy at zero inclination, and these parameters remain constant throughout our study. By setting a nominal bulk Reynolds number of 6000, we provide a detailed statistical characterisation of the turbulent flow.

Our findings illustrate distinct changes in the flow regime with varying filament inclination. At lower angles, the canopy remains dense and significantly influences the flow, conforming to a classical canopy-flow regime. However, as the inclination approaches $90^{\circ}$, the intra-canopy region progressively becomes shielded from the outer flow. Remarkably, at $90^{\circ}$ inclination, the flow drag reduces significantly, and the total drag becomes lower than that typically seen in an open, filament-free flow.

We document this transition from a canopy-dominated regime to a scenario where the canopy becomes largely sheltered from the outer turbulent flow, highlighting key alterations in intra-canopy dynamics as filament inclination increases. Our observations are substantiated by an analysis of the velocity spectra, providing deeper insight into the interactions between the canopy and the developing turbulent boundary layer.
\end{abstract}


\maketitle 
\section{Introduction}
\label{sec:intro}
Canopy flows refer to fluid flows interacting with a dense array of obstacles or elements—such as vegetation, hairs, or engineered posts—that protrude from a surface into the flow. These flows commonly arise in natural settings and play significant geophysical and biological roles \cite{raupach1981turbulence, ghisalberti2002mixing, lodish2008molecular, goodwyn2009waterproof}. Notably, in certain configurations, natural canopies have been observed to promote turbulent drag reduction (e.g.\ seal fur, \cite{itoh2006turbulent}) or to enhance flight performance (e.g.\ bird feathers, \cite{brucker2014influence}), thus motivating innovative flow-control strategies.

\begin{figure}
\centering
\includegraphics[width=0.3\textwidth]{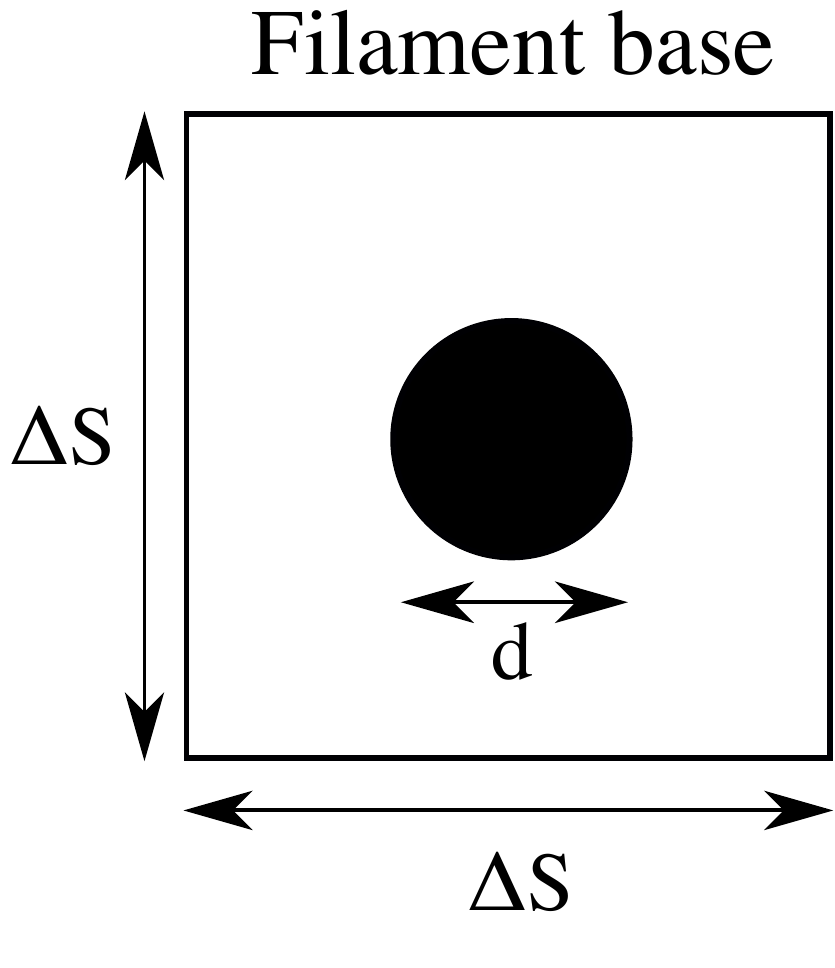}
\hfill
\includegraphics[width=0.42\textwidth]{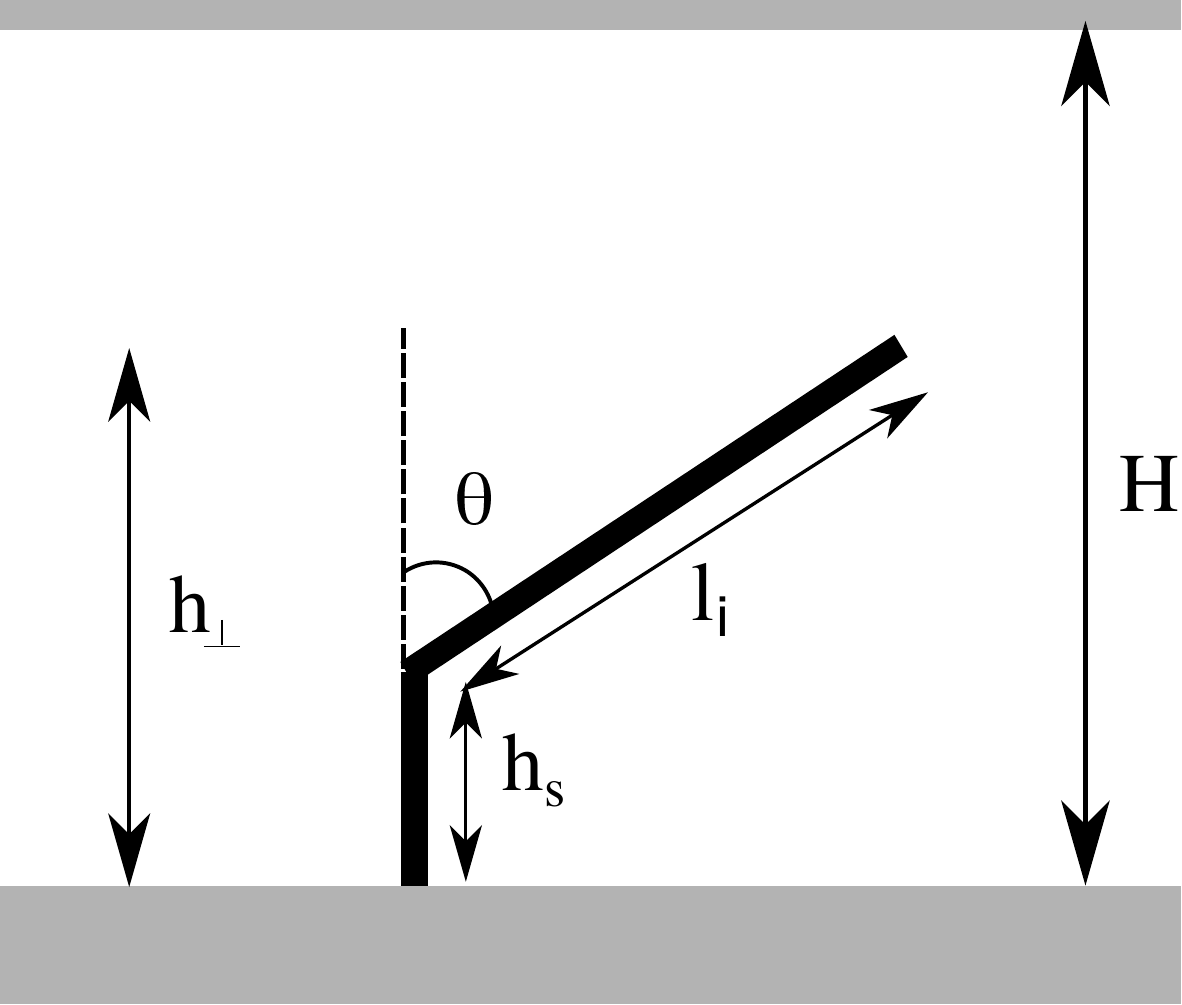}
\put(-305,-15){\large \textbf{(a)}}
\put(-100,-15){\large \textbf{(b)}}
\caption{Sketch of the geometrical parameters governing our inclined canopy, constituted by solid cylindrical filaments with diameter $d$ arranged in squared tiles of size $\Delta S$. 
Their shape is defined by the angle of inclination ($\theta$), the sheath region ($h_s$), the length of the inclined region ($l_i$) and the frontal projected height ($h$).}
\label{fig:intro}
\end{figure}

Such flows over canopies typically depend on geometric parameters, particularly the submergence (i.e.\ the ratio of canopy height $h$ to the flow depth $H$) and the solidity $\lambda$ \cite{nepf2012flow,monti2022solidity}. 
The solidity $\lambda$ is defined as the ratio between the frontal projected area of a canopy element and its base area (see Figure~\ref{fig:intro}):
\begin{equation}\label{eqlamba}
    \lambda=\int^{y=h}_{y=0}\frac{d(y)}{\Delta S^2}\,\mathrm{d}y,
\end{equation}
where $d(y)$ is the diameter of the canopy filaments as a function of the vertical coordinate $y$, and $\Delta S$ is their horizontal spacing. The submergence distinguishes between (i) emergent canopies, when $H/h \simeq 1$, and (ii) submerged canopies, if $H/h > 1$. This distinction highlights the extent of the boundary layer that develops above the canopy. Meanwhile, $\lambda$ characterises different flow regimes: for $\lambda \ll 0.1$ (sparse regime), the flow behaves like a turbulent boundary layer over distributed roughness \cite{poggi2004effect}, 
where turbulence reaches the substrate and the flow resembles a canonical rough-wall regime, although the canopy-induced form drag can still exceed the bed shear stress.
Conversely, for $\lambda \gg 0.1$ (dense regime), the filament form drag prevails.

Here, we focus on fully submerged and dense canopy configurations \cite{sharma2020scaling}, where the abrupt drag discontinuity at the filament tips typically triggers a Kelvin--Helmholtz (KH)-type instability \cite{finnigan2000turbulence}, leading to large-scale, spanwise-coherent structures that control momentum exchange within and above the canopy \cite{raupach1996coherent}. Analogous large-scale structures also appear in flows over permeable substrates \cite{jimenez2001turbulent, kuwata-suga-2017} and over certain ribleted surfaces \cite{garcia2011hydrodynamic}, although in those cases they are often associated 

Natural canopies commonly consist of flexible elements (e.g.\ seagrass, crops) that align with the dominant flow direction \cite{tschisgale-etal-2021, wang-etal-2022, monti-etal-2023}. Such streamlining reduces local drag by shielding the near-bed region \cite{luhar2011flow}, allowing for mutual sheltering \cite{raupach1991, monti2022solidity}. The shape of each filament can be described using a curvilinear abscissa $s$ (along the stem) and an inclination angle $\theta$ (relative to the wall-normal axis). \citet{chen2011flow} partition the filament into a lower “sheath” region, with $\theta \approx 0^\circ$ due to the clamping at the base, and an upper, inclined region, $\theta > 0^\circ$, where the filament becomes more streamlined. 

Recent studies have shown how flexible canopy elements undergo deformation in response to flow-induced loading, modifying the local inclination and generating substantial changes in drag and turbulence penetration \cite{rota2024dynamics, monti2023collective}. 
The present study can therefore be interpreted as a rigid analogue of such configurations, in which the filament inclination is prescribed rather than emerging from fluid--structure interaction. This modelling choice enables a systematic and controlled investigation of the hydrodynamic response to inclination alone, abstracting away the complexities associated with filament elasticity and deformation dynamics. While this simplification departs from the fully coupled fluid, structure interaction scenario, it reflects the steady-state configurations that flexible elements may attain under specific flow conditions. Moreover, there exist practical applications where rigid inclined roughness elements are intentionally designed to maintain fixed inclination angles across different regimes, including artificial vegetation, 3D-printed textured surfaces, passive flow-control coatings, or deployable morphing skins locked into place for optimal performance. In these cases, the inclined elements act effectively as static roughness geometries that replicate the drag-modulating effects of flexible canopies. Our results thus provide physical insight relevant both to the idealised understanding of inclined canopies and to the rational design of bio-inspired or engineered roughness elements.

Changes in filament configuration directly modify the canopy inclination relative to the wall-normal axis, thereby reducing the frontal projected height (see Figure~\ref{fig:intro}) \cite{vogel1984drag}. If $\theta$ increases, the effective canopy height and the apparent solidity $\lambda$ tend to decrease; under certain circumstances, this may induce a shift in flow regime (e.g.\ $\lambda < 0.15$, \cite{nicholas2022trans}). In such situations, which can be viewed as a local modulation of wall-normal permeability, simple models for transition (e.g.\ those introduced by \citet{monti2022solidity} based on eddy penetration depth) may be inadequate due to geometric constraints imposed by inclined filaments. Partial shielding may also spawn new flow regimes not yet explored in existing literature. Exploring these transitions contributes to the broader goal of leveraging canopy geometries as tools for turbulent drag control and momentum redistribution in engineering flows.

While several studies have examined flows over vertical or moderately inclined canopies \citep{nepf2012flow, monti2022solidity, nicholas2022trans}, a systematic investigation of the flow physics induced by highly inclined geometries is still lacking. Previous work has mostly focused on roughness functions, mean profiles, or vortex generation, with limited attention paid to unifying models capable of predicting drag variation or penetration depth as a function of filament inclination. Moreover, classical models for drag partitioning or momentum exchange often assume vertical roughness elements or homogeneous permeability, making them ill-suited to capture the anisotropic geometrical effects introduced by inclination.

Some insight can be drawn from the literature on riblet-inspired surfaces and anisotropic roughness \citep{garcia2011hydrodynamic}, which has shown that slip velocity, effective protrusion height, and lateral sheltering all influence drag reduction. However, these configurations are typically limited to two-dimensional grooves and do not directly generalise to three-dimensional filament arrangements with varying inclination.

The framework of virtual origins, developed to model rough-wall flows and canopy turbulence, has proved effective in predicting mean flow and turbulence profiles \citep{coceal2007structure, monti2019large, lopez2022structure}. Nonetheless, the application of this framework to inclined canopies has not been explored in depth, particularly in regimes where the frontal area and solidity decrease continuously with increasing inclination. The potential of combining multiple virtual origins (e.g.\ for mean velocity, fluctuations, and stresses) into a predictive model for drag change has also not been realised in the context of inclined canopy flows.

 The present study aims to bridge this gap by systematically investigating the hydrodynamic response of turbulent open-channel flows to rigid canopy elements with varying inclination. Using high-resolution large eddy simulations (LES), we explore a range of configurations spanning from fully vertical to fully horizontal filaments, all with identical volumetric blockage. We propose a unified framework based on the virtual origin concept, extended to inclined geometries, and introduce an algebraic model that accurately captures drag variation across all cases. This model incorporates both geometrical parameters (e.g.\ frontal height and solidity) and flow-based virtual origins (mean, turbulent, and transpiration). 
By clarifying how inclination modifies turbulence penetration, virtual origin location, and ultimately drag production, our study provides novel insights into canopy-flow interactions and offers a compact predictive tool applicable to both natural and engineered surfaces. The broader aim is to understand and rationalise how anisotropic surface structures alter momentum transfer, with potential applications in drag reduction, bio-inspired design, and roughness modeling.

The remainder of this manuscript is organised as follows. In the next section, we describe our numerical approach. In the results section, we present the statistical characterisation of all canopy configurations, accompanied by instantaneous flow visualisations. Finally, a conclusion section summarises the key outcomes of this work.
\section{Numerical Techniques}
\label{numerics}

We conduct our LES simulations using an in-house developed, incompressible Navier--Stokes solver called \textsc{SUSA} \cite{omidyeganeh2013large}, filtering the velocity and pressure fluctuations that occur below a spatial threshold with a typical length scale lying within the inertial sub-range of turbulence. Larger motions remain directly resolved.

In a Cartesian framework, the streamwise, wall-normal, and spanwise directions are represented by $x$, $y$, and $z$ (sometimes $x_1$, $x_2$, and $x_3$), while the corresponding velocity components are indicated by $u$, $v$, and $w$ (or $u_1$, $u_2$, and $u_3$), respectively. The dimensionless LES equations for the resolved fields (i.e.\ $\overline{u}$ and $\overline{p}$) thus read:
\begin{equation}\label{eq:cont}
    \frac{\partial \overline{u}_{i}}{\partial x_i} = 0,
\end{equation}
\begin{equation} \label{eq:momen}
    \frac{\partial \overline{u}_i}{\partial t}
    + \frac{\partial (\overline{u}_i\,\overline{u}_j)}{\partial x_j}
    = -\,\frac{\partial \overline{P}}{\partial x_i}
    + \frac{1}{Re_b}\,\frac{\partial^2 \overline{u}_i}{\partial x_j x_j}
    - \frac{\partial \tau_{i,j}}{\partial x_{j}}
    + f_i.
\end{equation}
In the formulation above, the bulk Reynolds number (based on the bulk velocity $U_b$, the open-channel height $H$, and the kinematic viscosity $\nu$) is defined as $Re_b = U_b H/\nu$, while $\tau_{i,j} = \overline{u_i u_j} - \overline{u_i}\,\overline{u_j}$ is the unresolved sub-grid Reynolds stress tensor \cite{leonard1975energy}. In particular, $\tau$ is modelled using an eddy-viscosity-based approach, known as the Integral Length Scale Approximation (ILSA) \cite{piomelli2015grid}. This method computes the model length scale and constant \textit{locally}, thus eliminating direct dependence on the underlying Eulerian grid \cite{rouhi2016dynamic}.  

The governing equations are spatially discretised with a second-order, cell-centred finite volume formulation. The pressure and velocity numerical grid nodes are collocated at the centre of each cell, and the appearance of spurious oscillations is avoided by deploying the deferred correction approach proposed by \citet{rhie1983numerical}. The LES equations are advanced in time using a second-order, semi-implicit fractional step method \cite{kim1985application}, where the wall-normal diffusion term is treated with a second-order, implicit Crank--Nicolson scheme, while an explicit Adams--Bashforth scheme is applied to the remaining terms. 

The Poisson equation for the pressure, which must be solved at each time step to satisfy the solenoidal condition for the velocity field, is transformed into a series of one-dimensional ($1$D) equations in wave-number space via an efficient multi-dimensional Fast Fourier Transform (FFT) \cite{dalcin2019fast} carried out along the wall-parallel directions. The resulting $1$D equations are then solved directly by means of a Cholesky factorisation technique \cite{higham2009cholesky}. The code is parallelised via the Message Passing Interface (MPI), using a domain decomposition approach. Further details on the code, its parallelisation, and the extensive validation campaign can be found in the literature \cite{omidyeganeh2013large,rosti2016direct}.

To discretise the canopy filaments, we adopt an Immersed Boundary Method (IBM) to impose the no-slip and no-penetration boundary conditions on each stem. Each filament is represented by a set of Lagrangian nodes distributed along its length, independently of the underlying Eulerian grid. Around each Lagrangian node, the IBM defines a compact support over which a spatially distributed set of body forces is applied to locally drive the fluid velocity to zero. This support corresponds to the effective domain of a regularised delta function (e.g., a 4-point kernel), spanning a small neighbourhood of grid cells. It enables smooth coupling between the solid structure and the surrounding fluid.

The size of each support is determined by the shape of the regularisation kernel and the local mesh resolution. Although the kernel typically distributes the forcing over four grid cells in each spatial direction, this support size does not directly correspond to the physical diameter of the filament. Instead, the hydrodynamic thickness (i.e. the effective diameter as perceived by the flow) must be inferred empirically through numerical calibration. Based on prior validation studies involving filaments mounted perpendicular to the wall (Favier et al. \cite{favier2014lattice}, Monti et al. \cite{monti2019large}), the effective filament diameter enforced by the IBM is estimated to be 
$d \simeq 2.2 \Delta x$ (or $d \simeq 2.2 \Delta z$, noting that $\Delta x=\Delta z$ in the present configuration). 
This estimate comes from direct comparisons of flow quantities, such as the drag coefficient $C_D$
against benchmark results for canonical flows (e.g. flow past a cylinder) using direct forcing methods \cite{monti2019large}.

Importantly, while the estimate $d \simeq 2.2 \Delta x$ for the hydrodynamic thickness is well validated for filaments mounted perpendicular to the wall, its applicability to inclined filaments requires careful interpretation. 
In inclined configurations, the Lagrangian points—although equispaced along the filament length—are no longer evenly spaced in the Cartesian directions. Since the regularisation kernel is aligned with the Eulerian grid, the volumetric forcing it produces remains isotropic in grid space and does not rotate with the filament. As a result, the support over which body forces are applied becomes misaligned with the filament’s local geometry, leading to mild geometric inconsistencies near the interface. Nevertheless, this mismatch has limited hydrodynamic impact, especially at higher inclination angles: as filaments become more aligned with the flow direction, the dominant contribution to drag transitions from pressure (form) drag to viscous (skin friction) drag. In this regime, the exact value of the hydrodynamic cross-sectional thickness becomes less critical. Consequently, the approximation $d \simeq 2.2 \Delta x$
emains a practical and robust estimate across a range of inclinations, particularly when evaluating global flow statistics rather than local force distributions.

More technical details on the IBM formulation and implementation are given in \cite{pinelli2010immersed}, while its application and validation in the context of filament-resolving canopy flows—including calibration of the support cage and analysis of force resolution—are thoroughly presented in \cite{monti2019large}. In the same reference, a region-by-region quantitative breakdown of sub-grid-scale effects is also provided.

Following the approach adopted in previous investigations by \citet{monti2020genesis}, \citet{nicholas2022}, and \citet{monti2022solidity}, the filaments are uniformly distributed across the impermeable channel wall by partitioning it into a regular array of non-overlapping square tiles of size $\Delta S$. Within each tile, the base of a single filament is placed at a location sampled from a uniform two-dimensional distribution confined to the tile's boundaries. This random placement strategy eliminates spatial periodicity in the canopy layout and mitigates the risk of preferential flow alignment or artificial channelling within the canopy layer, thereby preserving statistical isotropy in the wall-parallel directions.

Inspired by naturally occurring aquatic canopies, we consider an idealised set of rigid canopy configurations where the overall filament geometry is defined by a sheath and an inclined region. The former remains perpendicular to the impermeable wall, while the latter is governed by the angle of inclination ($\theta$), which is the free parameter of our investigation (see Figure~\ref{fig:intro}). The systematic variation of $\theta \in [0^\circ, 90^\circ]$ leads to a modulation of the frontal and vertical projected lengths of the filament, corresponding to a variation of the solidity $\lambda$ approximately within $[0.1, 0.5]$. All the other geometrical parameters are chosen so that the nearly upright configuration ($\theta = 0$) matches the fully dense regime ($\lambda \approx 0.56$) of \citet{nicholas2022}. The total length of the filaments is kept at $l = h_s + l_i = 0.1\,H$, with $h_s = 0.1\,l$ and $l_i = 0.9\,l$, while also enforcing a constant geometrical ratio $Hd/\Delta S^2 = 5.6$. 

\begin{table}[!t]
  \begin{center}
  \def~{\hphantom{0}}
  \begin{tabular}{|l|c|c|c|c|c|c|c|c|}
  \hline
  \textbf{Configuration} & \boldmath$\theta$ & \boldmath$h_{\bot}/H$ & \boldmath$\Delta S / h_{\bot}$ & \boldmath$Re_{\tau,o}$ &
  \boldmath$\Delta x_o^+$, \boldmath$\Delta z_o^+$ & \boldmath$\Delta y_{w,o}^+$, \boldmath$\Delta y_{h,o}^+$ & \textbf{Symbol}\\[3pt]
  \hline
  Negligible Inclination & 0 & 0.1 & 0.66 & 717 & 8.4 & 0.38 & $\nabla$\\
  \hline
  Weak Inclination       & 33.5 & 0.085 & 0.77 & 597 & 7   & 0.32 & $\square$\\
  \hline
  Mild Inclination       & 48.25 & 0.07 & 0.94 & 544 & 6.3 & 0.27 & $\triangle$\\
  \hline
  Relatively Strong Inclination & 60 & 0.055 & 1.19 & 478 & 5.47 & 0.25 & $+$\\
  \hline
  Strong Inclination    & 77.5 & 0.03 & 2.18 & 334 & 3.7 & 0.17 & o\\
  \hline
  Very Strong Inclination & 90 & 0.01 & 6.55 & 317 & 3.5 & 0.16 & \sixpointstar \ \\
  \hline
  Smooth Open Channel & -- & -- & -- & 326 & 3.6 & 0.16 & --\\
  \hline
  \end{tabular}
  \caption{Simulation parameters considered in our study. Filaments are distributed in $n_i \times n_k = 96 \times 72$ streamwise and spanwise rows, respectively. $Re_{\tau,o}$ is the friction Reynolds number based on the friction velocity ($u_{\tau,out}$) at the virtual origin. The grid spacing in the homogeneous directions is $\Delta x^+$ and $\Delta z^+$. The wall-normal grid spacing near the impermeable wall and at the canopy tip is $\Delta y_{w,o}^+$ and $\Delta y_{h,o}^+$.}
  \label{tab:canopy_simulation}
  \end{center}
\end{table}

We simulate all canopy configurations on a Cartesian mesh characterised by a uniform distribution of grid points along the horizontal (wall-parallel) directions, and a wall-normal spacing that is uniform within the canopy region, followed by a hyperbolic tangent stretching above the canopy tips up to the free-slip boundary at the top of the channel.

Similarly to the previous work by \citet{monti2020genesis}, all configurations share the same computational box of size $L_x/H = 2\pi$, $L_y/H = 1$, and $L_z/H = 1.5\pi$, but the streamwise length is doubled (i.e.\ $L_x/H=4\pi$) for the configurations with $\theta = 77^\circ, 90^\circ$ in order to capture large-scale coherent structures. The total number of grid points is $N_z=432$ and $N_x=576,\,1152$ for $\theta \in \{0^\circ, 30^\circ, 48.15^\circ, 60^\circ\},\,\{77^\circ, 90^\circ\}$, while the points in the wall-normal direction range within $N_y \in [216,260]$.

As a result, the grid resolution in wall units (see Table~I) satisfies the threshold needed to correctly simulate canonical wall-bounded turbulent flows \cite{kim1987turbulence}. That resolution is computed based on the outer friction velocity, which relates to the total stress evaluated at the virtual origin (i.e.\ $u_{\tau,out} = \sqrt{\nu\,(\partial \langle U \rangle/\partial y)_{vo} - \langle uv \rangle_{vo}}$). 
Here, the virtual origin \( y_{vo} \) refers to the effective wall-normal origin perceived by the outer flow, specifically, the position where the origin must be placed for the mean velocity profile to recover a canonical logarithmic shape~\cite{monti2020genesis}.

Further discussion on viscous scaling and the determination of the virtual origin is provided in the next section.

For reference, we have also simulated a smooth-wall turbulent open-channel flow, using a domain similar to that of the extremely inclined canopy cases. In that scenario, the numerical grid resolution in wall units (non-dimensionalised using the friction velocity at the wall, $u_{\tau,in} = \sqrt{\nu\,(\partial \langle U \rangle/\partial y)_w}$) again proves adequate, as documented in Table~\ref{tab:canopy_simulation}. More details on the extensive validation campaign can be found in previous works \cite{omidyeganeh2013large,rosti2016direct}.

Regarding boundary treatment, we enforce periodic conditions along the wall-parallel directions, no-slip at the bottom impermeable wall, and free slip at the top surface for all cases. The flow in the domain is driven by a uniform pressure gradient applied along the streamwise direction. At each time step, this pressure gradient is adjusted to obtain a time-independent, constant volumetric flow rate, corresponding to a bulk Reynolds number of $Re_b = U_b H/\nu = 6000$. This value matches that of \citet{monti2020genesis, nicholas2022, monti2022solidity}, enabling direct comparisons with different canopy configurations.

To facilitate the reader's understanding of the subsequent analysis, we provide in Table~II 
a summary of all length scales and associated dimensionless parameters used throughout the manuscript.

\renewcommand{\arraystretch}{0.8}
\setlength{\tabcolsep}{6pt}
\small
\begin{table}[htbp!]
\centering
\color{black}
\begin{tabular}{@{}l p{4.2cm} p{8.3cm}@{}}
\toprule
\textbf{Symbol} & \textbf{Definition} & \textbf{Physical Meaning} \\
\midrule
\multicolumn{3}{l}{\textbf{Geometric Length Scales}} \\
$H$ & Open channel height & Total height of the computational domain \\
$h$ & Canopy height & Height of perpendicular filaments \\
$h_\perp$ & Frontal projected height & $h_s + l_i \cos\theta$ -- height projected onto the flow direction \\
$h_s$ & Sheath region height & Height of the perpendicular base portion (10\% of total) \\
$l_i$ & Inclined region length & Length of the inclined portion of the filament \\
$l$ & Total filament length & $h_s + l_i = 0.1H$ (constant) \\
$\Delta S$ & Filament spacing & Distance between filament centres \\
$d$ & Filament diameter & Cross-sectional diameter of cylindrical filaments \\[0.3ex]
\midrule
\multicolumn{3}{l}{\textbf{Flow-Derived Length Scales}} \\
$y_{vo}$ & Virtual origin position & Abstract wall location seen by the outer flow \\
$y_{ip}$ & Internal inflection point & Wall-normal location of the inner inflection point \\
$h_{gp}$ & Geometrical penetration depth & $\Delta S/(2\tan\theta)$ -- constrained penetration scale \\
$k_{eff}$ & Effective canopy height & $h_\perp - y_{vo}$ -- portion of canopy seen by the outer flow \\[0.3ex]
\midrule
\multicolumn{3}{l}{\textbf{Virtual Origin Framework (Dimensionless)}} \\
$\ell_U^+$ & Mean flow virtual origin & As $y_{vo}$ in viscous units\\
$\ell_T^+$ & Turbulence virtual origin & Abstract wall seen by turbulent fluctuations \\
$\ell_u^+$ & Streamwise fluctuation origin & Virtual origin for the $u'$ velocity component \\
$\ell_v^+$ & Wall-normal fluctuation origin & Virtual origin for the $v'$ velocity component \\
$\ell_w^+$ & Spanwise fluctuation origin & Virtual origin for the $w'$ velocity component \\
$\ell_{uv}^+$ & Reynolds stress origin & Virtual origin for $\langle u'v' \rangle$ \\
$\ell_N^+$ & Transpirational virtual origin & $\ell_u^+ \ell_w^+ / \ell_v^+$ -- non-linear correction term \\[0.3ex]
\midrule
\multicolumn{3}{l}{\textbf{Dimensionless Parameters}} \\
$\lambda$ & Solidity & $h_\perp d / \Delta S^2$ -- frontal area ratio \\
$\Lambda_{eff}$ & Effective aspect ratio & $k_{eff} / \Delta S$ -- effective canopy characterisation \\
$l_g^+$ & Riblet parameter & $\sqrt{A_g^+}$ -- viscous-scaled groove cross-sectional area \\
\bottomrule
\end{tabular}
\label{tab:length_scales}
\caption{Summary of length scales and dimensionless parameters used in the manuscript}
\end{table}

Finally, to facilitate the reading and comprehension of the manuscript, we summarise here the conventions adopted for temporal, spatial, and ensemble averaging, as well as for the triple decomposition of flow quantities.
Time averaging is denoted by overbars (e.g., $\overline{u}$) and is performed over statistically stationary intervals once transients have decayed. Spatial averaging is indicated using angle brackets, $\langle \cdot \rangle$, and typically involves horizontal planes (i.e., $x$–$z$ averaging at constant $y$). When both operators are applied, the result is a double average denoted as $\langle \overline{u} \rangle$. Throughout the manuscript, we also employ a triple decomposition of the velocity field:
\begin{equation}
u(x,y,z,t) = \langle \overline{u}(y) \rangle + \overline{u}^\prime(x,y,z) + u^{\prime\prime}(x,y,z,t),
\end{equation}
where $\langle \overline{u}(y) \rangle$ is the mean profile, $\overline{u}^\prime$ represents the coherent spatial deviation from the mean, and $u^{\prime\prime}$ captures the remaining incoherent (turbulent) fluctuations. These definitions enable consistent interpretation of the profiles and fluctuations presented in the following sections.
\section{Results and Discussion}
\label{results}

In this section, we present the results obtained from our six statistically converged simulations. The main focus is on comparing the statistical quantities and identifying the structures in the turbulent flow field. We begin by analysing the mean velocity distributions and their characteristics, then proceed to examine higher-order statistics and turbulent coherent structures. Where relevant, we also support our discussion by using results from \citet{monti2020genesis} and \citet{nicholas2022}, shown by red and blue crossed lines/symbols, respectively.

\subsection{Statistical characterisation of the flow}

We start by plotting the mean velocity profile on a semi-logarithmic scale distinguishing
between two distinct layers: an inner layer, corresponding to the canopy bed, and an outer layer, associated with an abstract plane located at the virtual origin  $y_{vo}$ that typically lies below the canopy tip. The presence of an outwardly shifted boundary layer has been previously noted in studies on rough walls  \citep{jimenez2004turbulent}, permeable substrates  \citep{rosti2018},  and canopies \citep{monti2019large,nicholas2022trans}.   
Although the definition of the virtual origin is provided in the cited publications, we report here the methodology used for its determination for the sake of completeness. The position of the virtual origin, \( y_{vo} \), is obtained by enforcing the mean outer flow to follow a canonical logarithmic profile, i.e.,
\begin{equation}
\langle u \rangle = \frac{u_{\tau,\text{out}}}{\kappa} \log \left( \frac{(y - y_{vo}) u_{\tau,\text{out}}}{\nu} \right) + B.
\label{virtu}
\end{equation}
This expression is a standard modification of the boundary-layer log law for flows over rough surfaces~\citep{jimenez2004turbulent}. In Eq.~\eqref{virtu}, \( \kappa \) is the von Kármán constant, and \( u_{\tau,\text{out}} \) is the friction velocity computed from the total stress evaluated at the virtual origin \( y_{vo} \), i.e.
\begin{equation}
u_{\tau,\text{out}} = \left( \frac{\tau(y_{vo})}{\rho} \right)^{1/2} ~ \mbox{with} \; \; 
\tau(y_{vo}) = \mu \left. \frac{d \langle u \rangle}{dy} \right|_{y = y_{vo}} - \rho \langle u'v' \rangle(y_{vo}).
\end{equation}
If the total stress profile is known, the logarithmic law~\eqref{virtu} can be interpreted as an implicit equation for the unknown \( y_{vo} \), which is determined for each canopy configuration using the accumulated statistical values.
Table~\ref{virtut} reports the distance from the canopy tip to the virtual origin for all considered inclination angles.

\begin{table}[h]
\centering
{\color{black}
\begin{tabular}{@{}l|c|c|c|c|c|c|@{}}
\toprule
\(\theta\)                & 0     & 33.5  & 48.25 & 60    & 77.5  & 90    \\
\midrule
\(y_{\text{tip}}/H\)                & 0.105 & 0.085 & 0.072 & 0.055 & 0.030 & 0.015 \\
\midrule
\((y_{\text{tip}} - y_{vo})/H\)     & 0.030 & 0.015 & 0.012 & 0.005 & 0.000 & 0.000 \\
\bottomrule
\end{tabular}
\caption{Location of the canopy tip (\(y_{\text{tip}}/H\)) and distance to the virtual origin \((y_{\text{tip}} - y_{vo})/H\) for all inclination angles \(\theta\). In all cases, \( y_{vo} \) lies within the canopy.}
}
\label{virtut}
\end{table}

For the inner layer, we define an internal friction velocity (based on the wall shear stress at the canopy bed):
\begin{equation}
    u_{\tau,in} = \sqrt{\nu\,\bigl(\partial \langle U \rangle / \partial y\bigr)_w},
\end{equation}
For the outer layer,  an external friction velocity (based on the total stress at the virtual origin) is introduced as:
\begin{equation}
    u_{\tau,out} = \sqrt{\nu\,\bigl(\partial \langle U \rangle / \partial y\bigr)_{vo} \;-\;\langle u\,v\rangle_{vo}}.
\end{equation}

\begin{figure}
\centering
\includegraphics[width=0.7\textwidth]{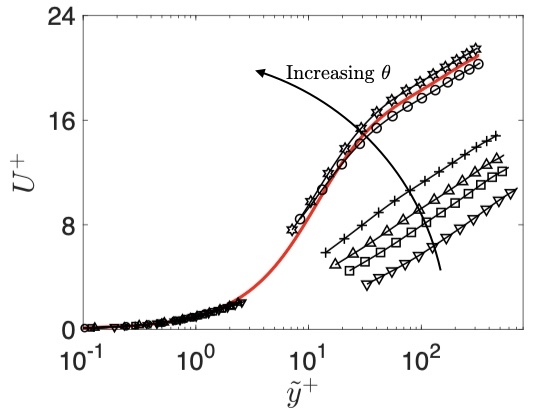}
\caption{
RMS velocity fluctuations and Reynolds stress profiles for all canopy configurations. 
Profiles are normalised using wall units: $u'$, $v'$, $w'$ and $\langle u'v' \rangle$ are scaled by $u_\tau$; $y^+$ is defined either from the wall-based friction velocity (i.e. $ u_{\tau,in}$) or from the total stress at the canopy tip ($ u_{\tau,out} $). 
This change in scaling definition introduces a small gap in the $y^+$ coordinate between approximately 3 and 8, visible across all cases.
Marker shapes indicate different inclination angles, as defined in Table~I, and colours follow a monotonic colormap (increasing with $\theta$) to help track trends.
}
\label{fig:loglaw}
\end{figure}

Figure~\ref{fig:loglaw} shows the mean velocity distributions scaled by these two friction velocities, also including the profile corresponding to a smooth-wall, turbulent, open-channel flow at $Re_{\tau}=326$. In the inner region (up to $\tilde{y}^+\approx5$, where $\tilde{y}^+$ is wall-normal coordinate scaled by the respective inner or outer  $u_{\tau}$), all canopy configurations closely match the smooth-wall profile, implying that the near-bed region is dominated by wall frictional drag, essentially independent of the canopy morphology (even in presence of the sheath layer of the filaments). Farther from the viscous sublayer, the filament inclination significantly affects the buffer layer, which is known to be independent from the outer flow \citep{jimenez1999autonomous}. In particular, the buffer-layer dynamics are modulated by the presence of inclined filaments that induce a meandering motion in the wall-parallel directions. 
As the inclination angle increases, the flow exhibits enhanced meandering within the canopy, a trend that will become evident in the section covering the flow structure.
For extremely inclined canopies ($\theta=77.5^\circ,\,90^\circ$), the frontal projected height in wall units (normalised by the outer friction velocity) ranges within $7 < h_{\bot}^+ < 12$, reminiscent of transitional or small-scale roughness \citep{thakkar2018direct}.  

In the logarithmic region (Figure~\ref{fig:loglaw}), the velocity distribution follows a universal log profile, highlighting its robustness. The log law for flow over textured surfaces \citep{jimenez2004turbulent} can be written as:
\begin{equation}\label{eq:log}
    U_{out}^+ = \frac{1}{\kappa} \ln(y_{out}^+) + B \;-\;\Delta U^+,
\end{equation}
where $y_{out} = y - y_{vo}$ is the shifted wall-normal coordinate, $\kappa=0.41$ is the usual von Kármán constant, and $B=5.5$ is the smooth-wall intercept. The subscript "out” denotes normalisation by $u_{\tau,out}$. The additional offset $\Delta U^+$ (the “resistance function”) captures the momentum deficit/surplus from the texture. This concept is common in rough-wall \citep{hama1954boundary,jimenez2004turbulent} and permeable substrates \citep{rosti2018,gomez2019turbulent}, where a positive $\Delta U^+$ indicates drag increase, while negative values suggest drag reduction. 
Across the six simulated canopy configurations, lower inclination angles $\theta$
exhibit the onset of a Kelvin–Helmholtz (KH)-like instability, which is typically associated with increased drag. In contrast, the flow topology in the highly inclined cases is markedly different, leading to drag levels comparable to, or even lower than, those of a smooth wall.

In particular, the value of $\Delta U^+$ decreases monotonically with increasing $\theta$, becoming negative for the most inclined filaments ($\theta=90^\circ$). Based on $\Delta U^+$, we can identify two asymptotic regimes: a canopy regime ($\theta\in\{0^\circ, 30^\circ, 48.25^\circ, 60^\circ\}$) with clear higher drag, and a \textit{roughness regime} ($\theta\in\{77.5^\circ, 90^\circ\}$) reminiscent of typical rough-wall boundary layers \citep{jimenez2004turbulent}. Between these extremes, there is a transition from drag increase to drag reduction near $\theta\approx84^\circ$, a value estimated by   linear interpolation of $\Delta U^+$.

\begin{figure}
\centering
\includegraphics[width=\textwidth]{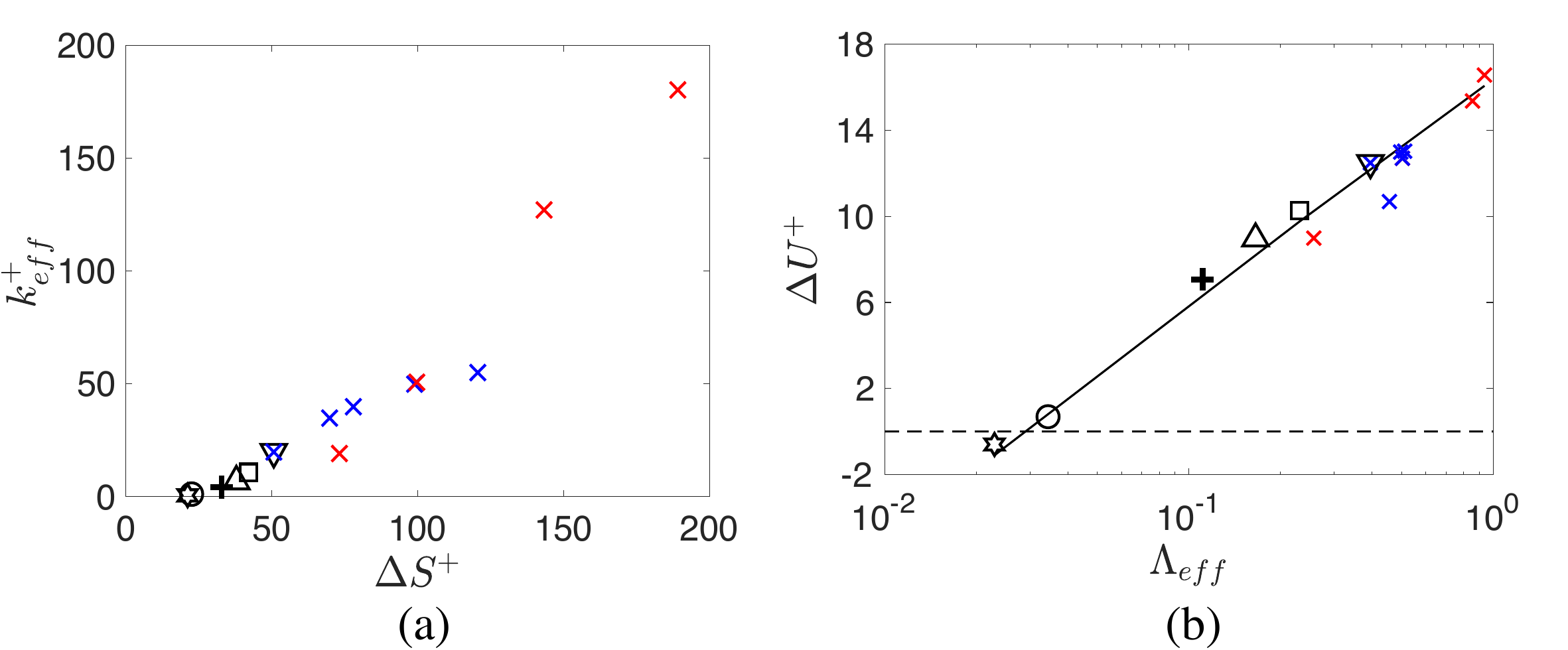}
\caption{(a) Variation of the effective canopy height $k_{\mathrm{eff}} = h_\perp - y_{vo}$ with the filament spacing $\Delta S$, illustrating the geometric control of the canopy penetration depth across inclination angles. (b) Corresponding variation of the drag offset $\Delta U^+$ with the effective aspect ratio $\Lambda_{\mathrm{eff}} = k_{\mathrm{eff}}/\Delta S$. The logarithmic trend in panel (b) aligns with classical roughness-function behaviour, while panel (a) highlights the direct geometric influence of spacing on the flow-perceived canopy height. Red and blue symbols represents the data extracted from \citet{monti2020genesis} and \citet{nicholas2022}, respectively.}
\label{fig:friction}
\end{figure}

Following \citet{monti2020genesis}, we define an effective canopy aspect ratio as 
\begin{equation}
\Lambda_{eff} = k_{eff} / \Delta S, 
\label{lambda_eff}
\end{equation}
where $ k_{eff} = h_{\bot} - y_{vo}$
denotes the depth of canopy penetrated by the outer-layer turbulence. Physically, $k_{eff}$ represents the portion of the canopy that actively interacts with the overlying eddies—typically smaller than the total frontal height of the filaments, especially when the inclination angle $\theta$ is large and the canopy becomes more streamlined along the flow direction.

On the other hand, the local filament spacing $\Delta S$ inside the canopy captures wall-normal permeability.

Figure~\ref{fig:friction} provides further insight into the role of the effective canopy height \(k_{eff}\) and its relation to drag modulation. Panel (a) shows that \(k_{\mathrm{eff}} = h_\perp - y_{vo}\) increases approximately linearly with filament spacing \(\Delta S\) for moderate to high inclinations. This trend reflects the geometric nature of canopy penetration depth as a function of spacing and inclination, independent of drag performance.
Panel (b) presents the variation of the drag offset \(\Delta U^+\) as a function of the effective aspect ratio \(\Lambda_{\mathrm{eff}} = k_{\mathrm{eff}}/\Delta S\). This normalised quantity is analogous to a roughness Reynolds number or dimensionless roughness height. As such, the logarithmic growth of \(\Delta U^+\) with \(\Lambda_{\mathrm{eff}}\) observed in Fig.~\ref{fig:friction}(b) is consistent with classical rough-wall behaviour and roughness-function correlations. The distinction between panels (a) and (b) is intentional: Fig.~\ref{fig:friction}(a) highlights the geometric control of the canopy configuration on the flow-perceived roughness, whereas Fig.~\ref{fig:friction}(b) relates that effective roughness to the induced drag modification.

\begin{figure}
\centering
\includegraphics[width=\textwidth]{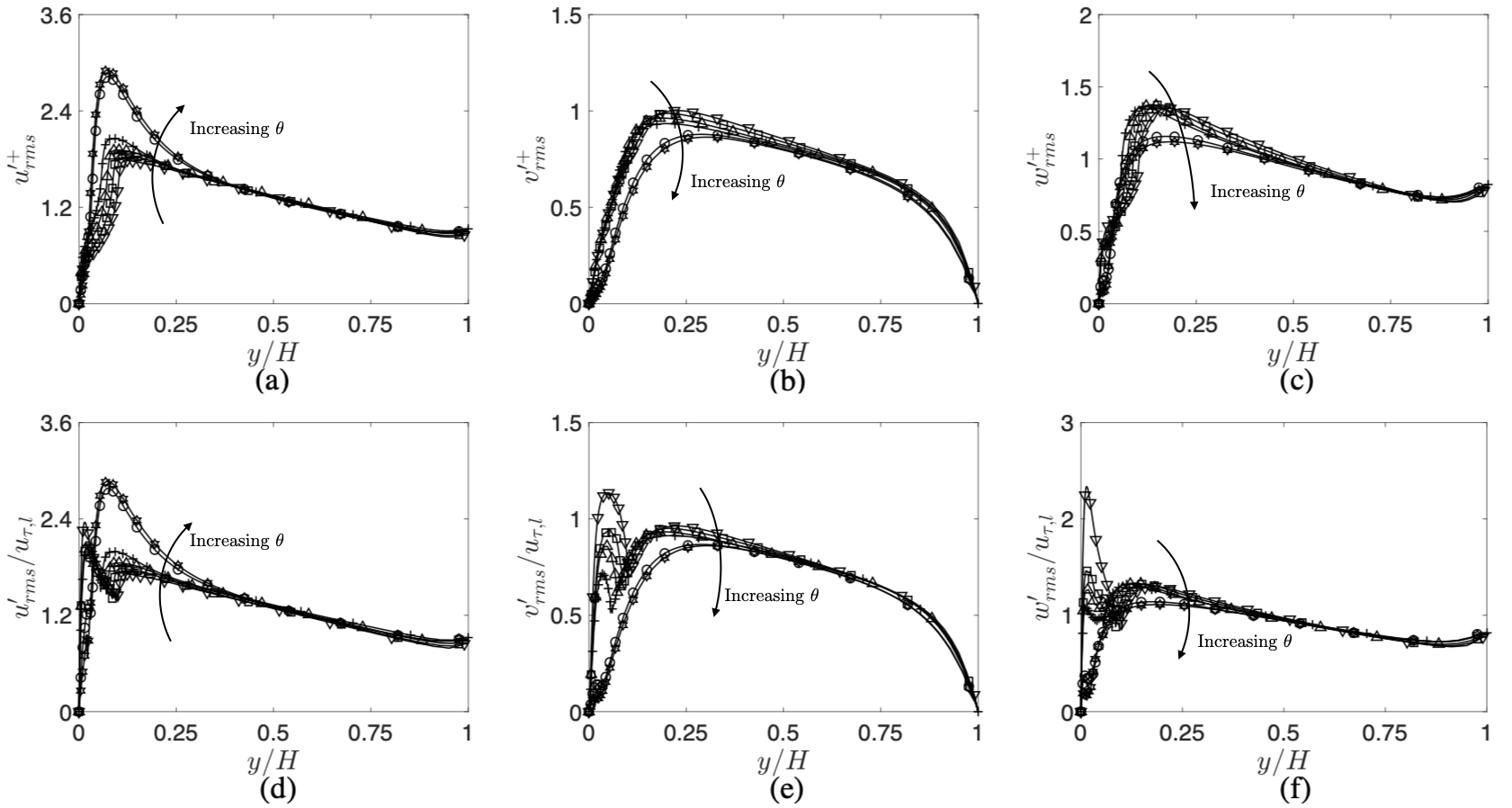}
\caption{
Root-mean-square (r.m.s.) velocity fluctuations of the streamwise ($u'$), wall-normal ($v'$), and spanwise ($w'$) components, normalised using both outer and local friction velocities (see bottom row), for all canopy configurations. 
Markers correspond to those defined in Table~I.
A curved arrow indicates the direction of increasing inclination angle $\theta$ across configurations.
}
\label{fig:fluct}
\end{figure}

We next examine velocity fluctuations, as they reveal how the filament inclination influences the turbulent flow. Figure~\ref{fig:fluct} shows the diagonal components of the Reynolds stress tensor vs. wall-normal distance. In the top row, these are scaled by $u_{\tau,out}$, and in the outer region, they collapse well, demonstrating Townsend’s outer-layer similarity \citep{townsend1976structure}. Consequently, the outer flow can be viewed as a rough boundary layer over a virtual wall \citep{monti2019large, scotti2006direct}. Within the canopy, however, $u_{\tau,out}$ scaling breaks down mainly because of the presence of the filaments. Following previous studies  \citep{sharma2020scaling, nicholas2022trans}, we select a local friction velocity: 
\begin{equation}\label{eq:locfricvelo}
    u_{\tau,l}=\sqrt{\frac{\nu\,\partial\langle U\rangle / \partial y \;-\;\langle u'v'\rangle}{1-y/H}},
\end{equation}
which balances viscous $\nu\partial\langle U\rangle / \partial y$ and Reynolds shear stress $\langle u'v'\rangle$ scaled by the non-dimensional distance $(1-y/H)$.
This scaling has been previously adopted in the context of sparse \citep{sharma2020scaling}, transitional \citep{nicholas2022trans} and dense \citep{monti2020genesis, nicholas2022} canopies, as well as manipulated shear flows \citep{jimenez2013near, tuerke2013simulations}. 

By observing the profiles normalised by the above formulation (see bottom row of Figure~\ref{fig:fluct}), we identify a smooth-wall-like behaviour with partial collapse, particularly in the outer region, enabling a meaningful comparison among the configurations. In the inner canopy region, some variability remains due to the inclination-dependent flow dynamics, which are discussed in detail later. The scaling presented by both friction velocities (i.e., outer and local) converges in the region that spans the external boundary layer developing above the canopy, owing to the absence of filaments in that region.

The distributions of 
 $\langle u'^2 \rangle^{1/2}$, $\langle v'^2 \rangle^{1/2}$, and $\langle w'^2 \rangle^{1/2}$
in Figure~\ref{fig:fluct} exhibit double peaks—one near the canopy bed and one above the canopy. For the streamwise velocity fluctuations, the outer peak reflects canonical streaky structures—alternating bands of high- and low-speed streamwise velocity—characteristic of wall-bounded turbulence and attributed to the action of streamwise vortices lifting and sweeping momentum \citep{finnigan2000turbulence, thakkar2018direct}. The outer peak increases in intensity with the increasing angle of inclination, suggesting a strengthening of the streaky structures. 

Conversely, the wall-normal and spanwise outer peaks decrease in intensity with increasing values of $\theta$. These trends can be connected to the characteristics of outer quasi-streamwise vortices (i.e. vortical structures whose axes are predominantly aligned with the streamwise direction and which typically reside above the canopy layer). As suggested by \citet{jimenez2013near}, the decrease in peak intensity may be associated with the weakening of these vortices due to reduced turbulence penetration from the outer flow into the canopy layer as inclination increases.

Differently, within the internal layer, we observe that the peak of the streamwise velocity fluctuations
decreases when the filament inclination is increased.
The presence of the peak deep within the canopy is linked to the structures generated by the flow around bi-periodic cylindrical filaments, driven by wall-normal jets \citep{monti2022solidity}. These jets emerge as the filaments disrupt the near-wall regenerative cycle and suppress the formation of streaky structures \citep{jimenez1999autonomous}. This behaviour suggests that the internal peak arises from the horizontal “squeezing” of the flow between the filament sheaths. This interpretation is further supported by the intra-canopy distributions of the spanwise velocity fluctuations, which exhibit patterns closely resembling those of the streamwise component. Notably, for the case with $\theta = 0^{\circ}$, the internal peaks of both the streamwise and spanwise velocity fluctuations show nearly identical intensities.

The internal peak of $v_{rms}$, similarly to the wall parallel components, decreases in intensity with increasing angles of inclination: this is indicative of the decrease in the level of penetration of the external flow due to the increasing filament inclination. 
For the extremely inclined configurations, the internal distributions for all three components do not feature an internal peak, presenting a departure from the behaviour exhibited by the low and mildly inclined configurations. Instead, the overall behaviour of the profiles for the roughness regime is almost smooth wall like \citep{moser1999direct}. 
Further insight into the modulation presented by the configurations at $\theta=77.5^{\circ}, 90^{\circ}$, along with their comparison with the canonical turbulent channel flow, are provided in the following subsection. 

\subsection{Virtual origin framework in the roughness regime}
\label{drregime}

Here, we focus on the roughness regime seen in inclined canopies whose drag becomes comparable to that of a smooth wall. These extremely inclined canopies behave like small-texture surfaces ($h_{\bot}^+<10$), influencing near-wall flow primarily by restricting cross-flow, while leaving the streamwise velocity component largely unaffected \citep{luchini1991resistance}.
This mechanism underpins several passive drag-reduction techniques (riblets \citep{bechert1997experiments}, superhydrophobic surfaces  \citep{rothstein2010slip, rosti2018}, etc.) and can be understood by introducing a “virtual origin” approach \citep{ibrahim2021smooth}, wherein turbulence effectively perceives a shifted wall location.

Following this approach we adopt a unifying framework that enables direct comparisons with smooth-wall turbulent flows. To remain consistent with studies on small textured surfaces, we define the resistance function $\Delta U^+$ as the change in drag relative to a smooth-wall turbulent open-channel flow simulated at the same nominal bulk Reynolds number (see the \textit{Numerical Techniques} section).

For the extremely inclined configurations considered in this study, we introduce an alternative wall-normal coordinate with its origin located at the tip of the filament, defined as $y_{h_{\bot}} = y - h_{\bot}$. In this coordinate system, any virtual origin lies below the canopy tip, i.e., $y_{h_{\bot}}^+ = -\ell_U^+$ or $y_{h_{\bot}}^+ = -\ell_T^+$. The virtual origin represents the location of an equivalent smooth wall as perceived by the turbulence (i.e., $\ell_T^+ = h_{\bot}^+ - y_{vo}^+$), inferred based on a fit to the canonical logarithmic law.

\begin{figure}
\centering
\includegraphics[width=\textwidth]{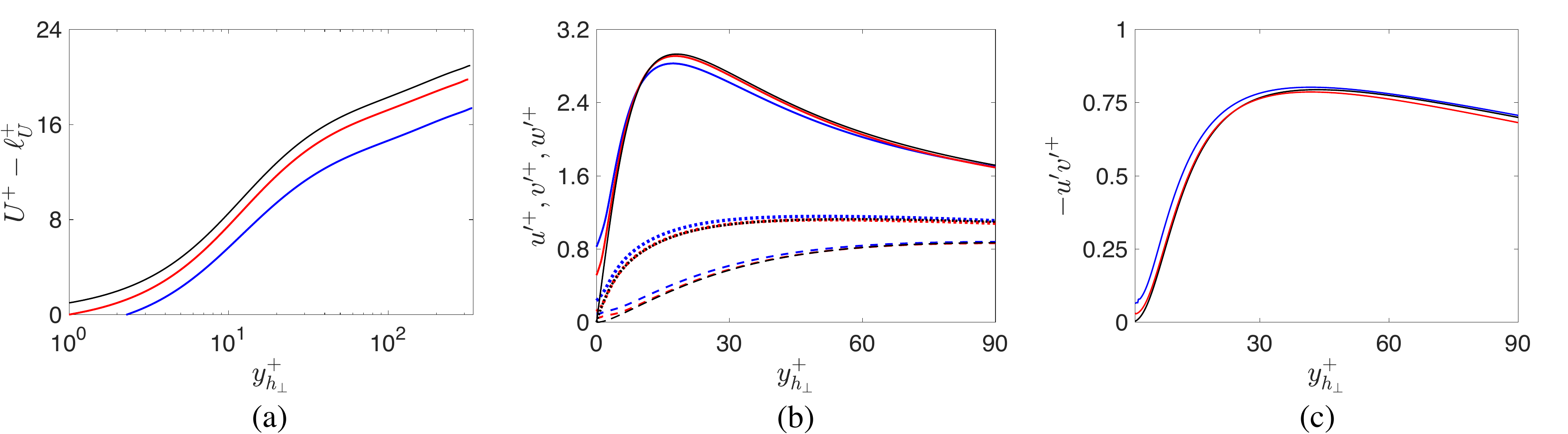}
\caption{Conventional approach for turbulence statistics with the origin at the canopy tip: (a) mean velocity, (b) r.m.s.\ velocity fluctuations, (c) Reynolds shear stress. Black, blue, and red curves represent $Re_{\tau}=326$ smooth wall, $\theta=77.5^\circ$, and $\theta=90^\circ$, respectively. Solid, dashed, and dotted lines (in b,c) are the streamwise, wall-normal, and spanwise components.}
\label{fig:drconventional}
\end{figure}

\citet{ibrahim2021smooth} argued that the virtual origin corresponding to the Reynolds shear stress provides a more appropriate reference for the origin of turbulence, defined as $\ell_T^+ = \ell_{uv}^+$. This perspective is rooted in the mean momentum balance for smooth-wall flows, where the dominant contributions come from viscous and Reynolds shear stresses, and where $\langle u'v'\rangle^+ = 0$ at the wall. Consequently, in flows that retain smooth-wall-like characteristics under a turbulence displacement, the appropriate origin is where $\langle u'v'\rangle^+$ perceives a smooth wall.

To evaluate both approaches, we determine the virtual origin for turbulence based on a best-fit shift of the Reynolds shear stress profile to smooth-wall data within the near-wall region ($10 < y_{h_{\bot}}^+ < 25$), following the method of \citet{ibrahim2021smooth}. This yields $\ell_{uv}^+ = 1.2$ and $\ell_{uv}^+ = 0.5$ for the $\theta = 77.5^\circ$ and $\theta = 90^\circ$ configurations respectively, both well within the threshold of $\ell_T^+ \leq 5$ proposed by \citet{garcia2019control}.

\begin{figure}
\centering
\includegraphics[width=\textwidth]{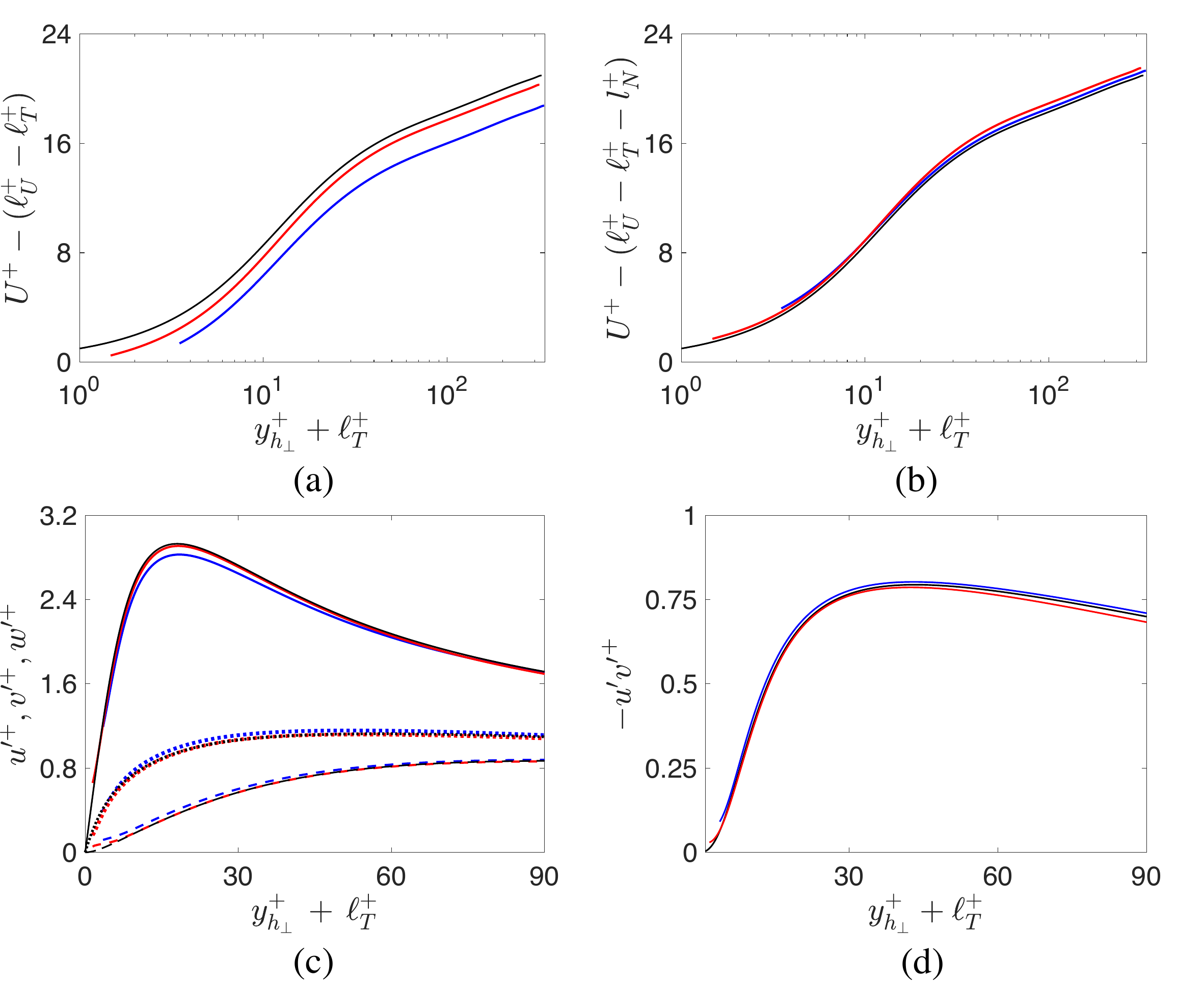}
\caption{Turbulence statistics shifted by the turbulence origin $\ell_T^+$: (a) mean velocity (shifted by $\Delta U^+$), (b) r.m.s.\ velocity fluctuations, (c) Reynolds shear stress. Black, blue, and red again represent the smooth wall at $Re_{\tau}=326$, $\theta=77.5^\circ$, and $\theta=90^\circ$. Solid/dashed/dotted lines in (b) are $u'$, $v'$, and $w'$.}
\label{fig:dreduction}
\end{figure}

The conventional approach for presenting turbulence statistics in the context of passive flow control techniques—such as riblets or superhydrophobic surfaces—defines the wall-normal origin at the crest of the surface textures (i.e., $y^+ = 0$) \citep{garcia2019control}. Following this convention, in Figure~\ref{fig:drconventional}, we present the mean velocity profiles by subtracting $\ell_U^+$, the slip velocity at the canopy tip, under the assumption of a logarithmic velocity distribution. 

This representation reveals distinct differences compared to smooth-wall data. As shown in Figures~\ref{fig:drconventional}b and \ref{fig:drconventional}c, the profiles of velocity fluctuations and Reynolds shear stress are shifted towards $y_{h_{\bot}} = 0$, suggesting that the near-wall turbulence structure is modified or dampened by the presence of inclined canopy elements. Consequently, the turbulence is no longer directly comparable to that of a canonical smooth-wall case.

However, the virtual origin framework proposed by \citet{ibrahim2021smooth} provides a physical interpretation for these observations. By redefining the wall-normal coordinate relative to the virtual origin of turbulence, such that $y_{h_{\bot}}^+ = -\ell_T^+$, the turbulence statistics can be reconciled with smooth-wall behaviour. Figures~\ref{fig:dreduction}a, \ref{fig:dreduction}c, and \ref{fig:dreduction}d show the mean velocity, velocity fluctuations, and Reynolds shear stress plotted against the modified wall-normal coordinate. Aside from the mean velocity, which remains slightly affected, the other quantities exhibit excellent agreement with the smooth-wall reference data. This confirms that, despite geometric alterations, the turbulence structure in the highly inclined canopy cases remains essentially unmodified, and the offset $\ell_T^+$ effectively accounts for the perceived origin shift in near-wall dynamics \citep{garcia2019control}.

The slight discrepancies observed in the mean velocity suggest that the drag change may be marginally underestimated. This mismatch can be attributed to a breakdown of the drag change law beyond its nominal range of validity (i.e., $\ell_T^+ \le 5$). 

To address this limitation, we introduce a corrective modification:
\begin{equation}
\Delta U^+ = \ell_U^+ - \ell_T^+ - \ell_N^+,
\label{eq:deltau}
\end{equation}
The first two terms correspond to the classical virtual origin formulation, while 
the third term, $\ell_N^+ = \ell_u^+ \ell_w^+ / \ell_v^+$, provides a mathematical 
correction that accounts for the enhanced drag in regimes where the original framework 
breaks down ($\ell_T^+ > 5$). 
While this correction enables accurate drag prediction 
across all configurations, we emphasize that the transpiration velocity $v_h'^+$ 
(shown later in Figure~9d) represents the fundamental physical scale governing drag 
changes. 
Here, we use the term \textit{transpiration} in the context of wall-bounded turbulence to denote the wall-normal mass flux across the canopy interface—distinct from its usage in terrestrial canopy literature, where it typically refers to biological processes. This velocity provides a more direct measure of transpiration intensity than the derived quantity $\ell_N^+$.

To evaluate $\ell_N^+$, we compute the virtual origins $\ell_u^+$, $\ell_v^+$, 
and $\ell_w^+$ using the same methodology employed to determine the virtual origin 
of the Reynolds shear stress. Specifically, each virtual origin is determined by 
finding the optimal vertical shift required to collapse the respective velocity 
component profile with smooth-wall reference data within the near-wall region 
($10 < y^+_{h_\perp} < 25$), where the superscript $+$ denotes normalization by 
the consistent friction velocity $u_{\tau,\text{out}}$ for all components. The 
optimal shift is determined by minimizing the root-mean-square deviation between 
the shifted canopy profile and the smooth-wall reference.

\begin{figure}
\centering
\includegraphics[width=\textwidth]{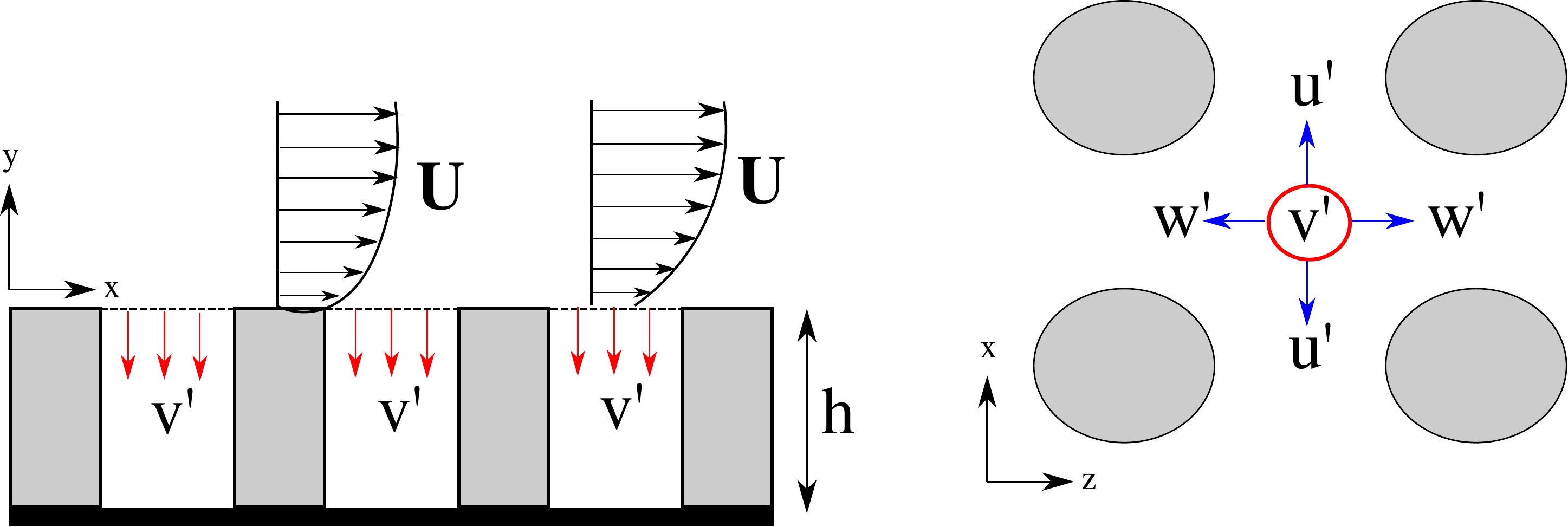}
\put(-290,-15){\large \textbf{(a)}}
\put(-70,-15){\large \textbf{(b)}}
\caption{(a) Side view illustrating the mean velocity variation along $x$, inducing a transpiration through the filament layer; (b) top view showing how momentum influx is redistributed in the wall-parallel directions.}
\label{fig:sketch}
\end{figure}

From a physical perspective, the streamwise and spanwise slip at the tip of the canopy induces a flux through vertical planes oriented along the streamwise direction \cite{khorasani2022near}, that is, planes extending in the wall-normal direction and aligned with the main flow.
According to mass conservation, these fluxes must be balanced by wall-normal transpiration through the horizontal plane at the canopy tip. A similar phenomenon has been observed in canopy flows, where the transpiration manifests as wall-normal jets impinging on the canopy bed \citep{monti2020genesis,monti2022solidity,nicholas2022}. To satisfy momentum conservation and the solenoidal condition, these jets deflect laterally along the wall-parallel directions, as illustrated conceptually in Figure~\ref{fig:sketch}, resulting in a redistribution of momentum within the canopy.

The transpirational term $\ell_N^+$ thus captures the interplay between this vertical momentum injection and its redirection along the canopy plane. To evaluate $\ell_N^+$, we compute the virtual origins $\ell_u^+$, $\ell_w^+$, and $\ell_v^+$ using the same methodology employed to determine the virtual origin of the Reynolds shear stress. Typically, the virtual origins associated with $u'$ and $w'$ correspond to their effective slip lengths \citep{garcia2019control}. 

However, due to the complex nature of the $v'$ root-mean-square profile, $\ell_v^+$ is determined by identifying the virtual origin that provides the best collapse of the wall-normal fluctuation intensity in the inner canopy region. This is achieved by shifting each $v'^+$ profile vertically by a candidate $\ell_v^+$ and evaluating the residual variance across all configurations within a selected fitting window, typically $0 < y^+ - \ell_v^+ < 40$. The optimal $\ell_v^+$ is then defined as the value that minimises this variance, effectively extrapolating the apparent origin of wall-normal fluctuations. This approach is analogous to identifying a transpiration or wall-normal slip length. Collectively, these estimates define the non-linear contribution $\ell_N^+$, which encapsulates both the vertical flux at the canopy tip and its subsequent redistribution across the canopy surface.

\begin{figure}
\centering
\includegraphics[width=\textwidth]{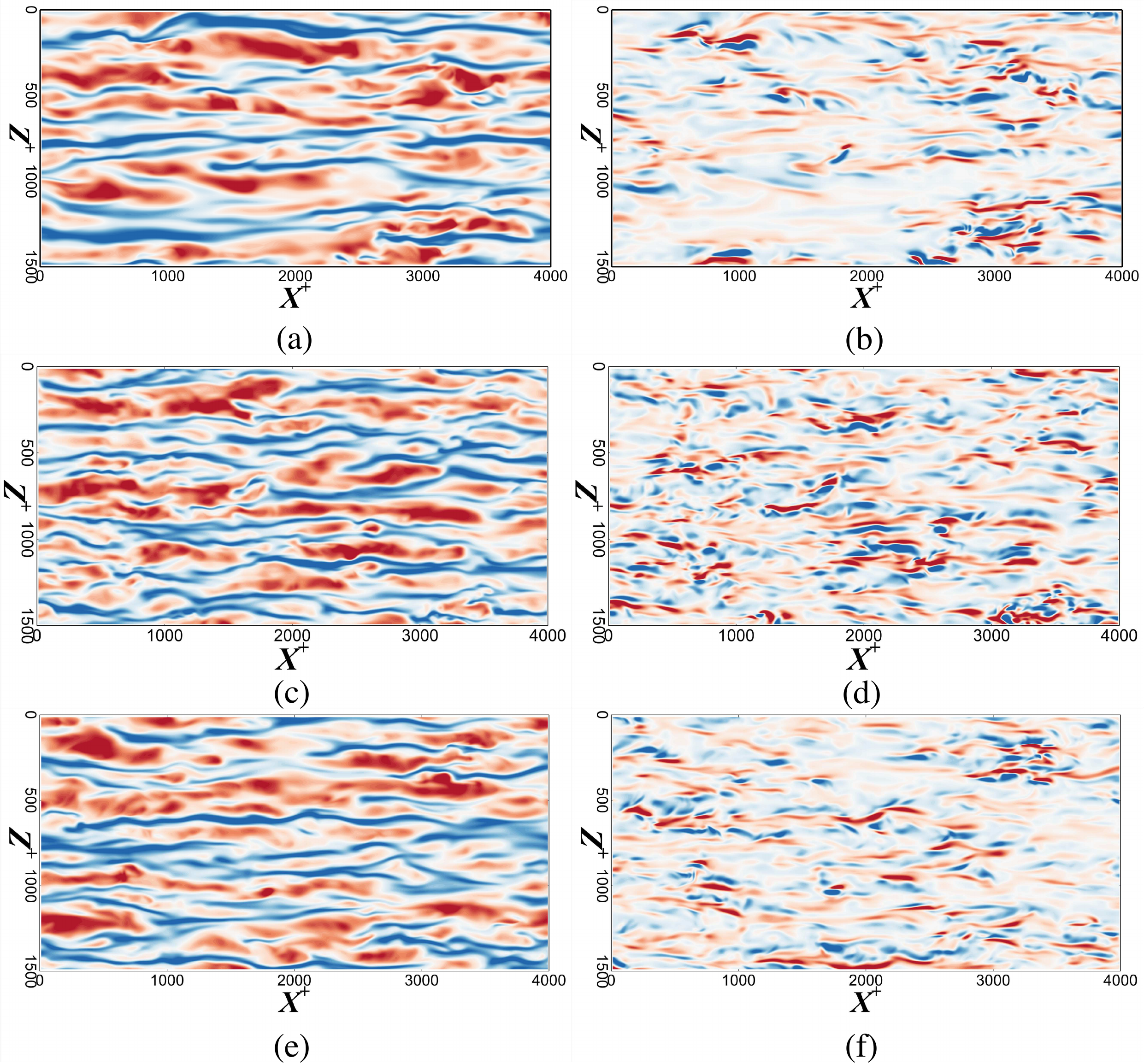}
\caption{Instantaneous realisations of the streamwise (left column) and wall-normal (right column) velocity fluctuations at $y_{h_{\bot}}^+ + \ell_T^+ = 20$. Rows (top to bottom): smooth-wall $Re_\tau=326$, $\theta=77.5^\circ$, and $\theta=90^\circ$. Colour bars scale with $u_{\tau,out}$ and $\nu$, ranging over $u'^{+}\in[-5,5]$ and $v'^{+}\in[-1,1]$.}
\label{fig:dr_snap}
\end{figure}

In Figure \ref{fig:dreduction}b, we present the mean velocity profile shifted using the modified drag change formulation, observing an excellent collapse with the smooth-wall data. This confirms that the drag change can be accurately captured by accounting for both the virtual origin of turbulence and the effects of transpiration. We hypothesise that the influence of extremely inclined filaments resembles that of small-scale surface textures, namely, a vertical displacement of the near-wall turbulence cycle rather than a fundamental alteration of its structure. To verify this, we examine instantaneous snapshots of the inclined canopy flows and the smooth-wall case. 

Figure~\ref{fig:dr_snap} compares the streamwise and wall-normal velocity fluctuations, $u^{\prime+}$ and $v^{\prime+}$, respectively, for the inclined canopies and the smooth open-channel flow at $y_{h_{\bot}}^+ + \ell_T^+ = 20$. 
This location corresponds to a constant offset above the frontal tip, aligned with the virtual origin $\ell_T^+$, allowing for consistent visualisation across configurations.

The visual similarity across the cases supports our hypothesis that the turbulence structure remains largely unmodified by the canopy and is comparable to that of a canonical smooth-wall flow. In particular, the inclined canopies do not appear to suppress or significantly deform the characteristic streaky patterns or wall-normal ejection/sweep events found in the smooth-wall case. 

In all configurations, elongated low- and high-speed streaks can be observed in the $u^{\prime+}$ snapshots, indicating the presence of coherent streamwise structures reminiscent of near-wall turbulence. Similarly, the $v^{\prime+}$ snapshots exhibit alternating regions of upward and downward motions, consistent with quasi-streamwise vortices and wall-normal momentum exchange.
The fact that these features persist in the presence of inclined filaments, despite changes in form drag and turbulent penetration, suggests that the outer-layer dynamics remain largely governed by the same mechanisms as in smooth-wall turbulence.

\subsection{Prediction of drag change via the virtual origins}

In this subsection, we extend the virtual origin framework to the canopy regime, which exhibits a substantial deviation from smooth-wall-like behaviour. In this regime, the filamentous layer introduces additional drag-increasing mechanisms, making it unsuitable to describe the flow solely through the virtual origin concept proposed by \citet{ibrahim2021smooth}. The core assumption underlying the virtual origin framework is that the r.m.s. distributions of the diagonal Reynolds stress components remain similar to those in a smooth-wall flow—an indication that the near-wall turbulence cycle remains largely intact. According to \citet{ibrahim2021smooth}, this assumption holds provided the turbulence virtual origin satisfies $\ell_T^+ \leq 5$. Beyond this threshold, the fundamental assumptions begin to break down, as the characteristic size of the surface texture extends beyond the viscous sublayer \citep{pope2000turbulent}. 

This condition corresponds to what is commonly referred to as the transitionally rough regime, which is marked by a complex interplay between viscous and pressure drag mechanisms \citep{flack2018moving,thakkar2018direct}. As the surface roughness becomes even more pronounced, the flow eventually enters the fully rough regime, where drag is almost entirely dominated by pressure forces \citep{jimenez2004turbulent}. In our study, all configurations, except the one with $\theta = 60^\circ$, exceed the $\ell_T^+ \leq 5$ threshold and therefore lie outside the suggested bounds of the virtual origin model. 

Here, our goal is not to demonstrate that the canopy regime is dynamically equivalent to a smooth-wall flow. Rather, we seek to leverage the virtual origins of the fluctuating velocity components to predict the drag change induced by the canopy. We hypothesise that by incorporating a non-linear correction term, as previously discussed, the virtual origin framework can be meaningfully extended into regimes where $\ell_T^+ > 5$.

\begin{table}[!t]
  \centering
  \def~{\hphantom{0}}
  \begin{tabular}{|l|c|c|c|c|c|c|c|}
  \hline
   \textbf{Nominal} & \boldmath$\theta$ & \boldmath$U_s^+$ & \boldmath$\ell_T^+$ & \boldmath$\ell_u^+$ & \boldmath$\ell_v^+$ & \boldmath$\ell_w^+$ & \boldmath$\Delta U^+$ \\[6pt]
   \hline
   Negligible Inc.  & 0 & 1.9  & 20.3 & 12.9 & 77.4 & 22.2 & 12.5 \\
   \hline
   Weak Inc.        & 33.5 & 2.3 & 9.9  & 8.6  & 43.8 & 21.1 & 10.3 \\
   \hline
   Mild Inc.        & 48.3 & 2.5 & 6.4  & 8.0  & 31.6 & 19.1 & 9.0 \\
   \hline
   Rel. Strong Inc. & 60   & 2.7 & 3.8  & 6.9  & 19.6 & 11.6 & 7.1 \\
   \hline
   Strong Inc.      & 77.5 & 3.0 & 1.2  & 3.6  & 5.6  & 3.8  & 0.7  \\
   \hline
   Very Strong Inc. & 90   & 1.7 & 0.5  & 1.7  & 1.7  & 1.3  & -0.6 \\
   \hline
  \end{tabular}
  \caption{Parameters used in the virtual origin framework \citep{ibrahim2021smooth}. $U_s^+$ is the slip velocity at the canopy tip, $\ell_T^+$ is the turbulence origin, $\ell_u^+$, $\ell_v^+$, and $\ell_w^+$ are the velocity fluctuation origins, and $\Delta U^+$ is computed from Equation~\ref{eq:deltau}.}
  \label{tab:virtual}
\end{table}

We begin by computing the virtual origin for turbulence ($\ell_T^+$) and those associated with the fluctuating velocity components ($\ell_u^+$, $\ell_v^+$, and $\ell_w^+$) for all canopy configurations, following the methodology outlined in the previous subsection. The resulting values are reported in Table~\ref{tab:virtual}. As previously observed, the turbulence origin $\ell_T^+$ aligns closely with the values obtained from the logarithmic mean velocity fit, reaffirming the validity of using the log-law formulation to define the turbulence origin—even though it includes a mean velocity term.

With these values, the non-linear transpirational virtual origin $\ell_N^+$ can be determined. Table~\ref{tab:virtual} shows that the non-dimensional virtual origins increase as the filament inclination angle grows. This trend suggests that the outer flow in low and mildly inclined canopy configurations perceives a deeper virtual origin. By extension, the origin for the mean streamwise flow should exhibit a similar behaviour.

However, for the nearly upright configuration ($\theta = 0^\circ$), we observe a significant discrepancy between the origin for the mean flow and that for turbulence. In this case, the mean flow origin derived from the slip velocity at the canopy tip ($\ell_U^+ = U_s^+$) becomes negligible when compared to $\ell_T^+$ and the virtual origins of the fluctuating velocity components. This inconsistency implies that, for such configurations, the streamwise velocity gradient near the wall deviates significantly from unity in wall units, indicating a breakdown of the classical linear behaviour of the near-wall profile \citep{luchini1996reducing,garcia2019control}.

Accordingly, for cases where $U_s^+ \ll \ell_T^+$, it becomes necessary to redefine the origin for the streamwise flow based on the virtual origin of the streamwise velocity fluctuations, i.e., using $\ell_u^+$ instead of $\ell_U^+$. 
We stress that the introduction of $\ell_N^+$ is not based on a physical model but serves as a mathematical closure term, particularly useful when the slip-velocity-based origin $\ell_U^+$ becomes insufficient, such as in configurations where $U_s^+ \ll \ell_T^+$ and the interaction between canopy-induced fluctuations and the outer flow dominates.

It should be mentioned that the values of $\Delta U^+$ reported in Table~\ref{tab:virtual} do not match exactly the predictions obtained using the formal expression in Equation~(\ref{eq:deltau}). While this equation, i.e. 
\(\Delta U^+ = \ell_U^+ - \ell_T^+ - \ell_N^+,
\)
offers a useful framework for interpreting drag changes through virtual origin displacements, it provides only an approximate estimate of the drag offset observed in the DNS.
The small discrepancies originate primarily from uncertainties in the determination of the turbulence origin $\ell_T^+$, which is extracted from fluctuating fields exhibiting substantial spatial variability. Additionally, the non-linear correction term $\ell_N^+ = \ell_u^+ \ell_w^+ / \ell_v^+$ depends on three independently measured quantities, introducing compounding sensitivity. The assumption that $\ell_U^+ = U_s^+$, while physically motivated, may also contribute to slight deviations, especially for intermediate inclinations where the slip velocity is harder to characterise.
Despite these limitations, Equation~(\ref{eq:deltau}) captures well the overall trends in $\Delta U^+$ across the range of inclination angles, supporting its utility as a predictive and interpretative tool within the virtual origin framework.

\begin{figure}
\centering
\includegraphics[width=\textwidth]{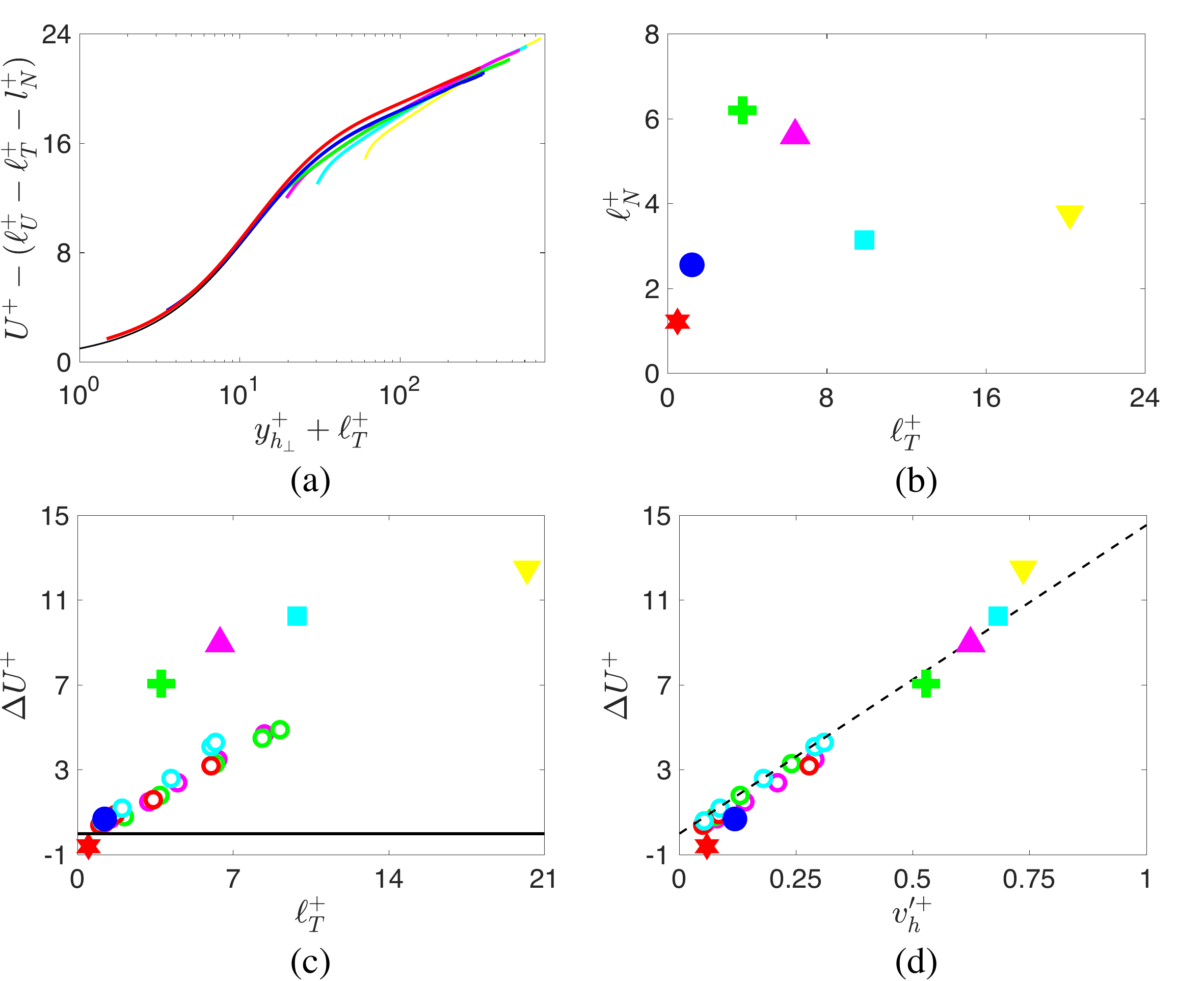}
\caption{(a) Mean velocity profile according to the drag change proposed in Equation~\ref{eq:deltau};  
(b) transpirational virtual origin as a function of the origin for the turbulence;  
(c) drag change as a function of the abstract origin for the turbulence; and  
(d) drag change as a function of the transpirational velocity at the tip of the canopy.  
The colours of the solid lines correspond to the inclination angles of the present simulations, as listed in Table~\ref{tab:canopy_simulation}.  
Coloured symbols represent data from \citet{abderrahaman2019modulation:  }
\textcolor{magenta}{O} — colocated posts;  
\textcolor{green}{O} — colocated posts with varying heights;  
\textcolor{red}{O} — staggered posts (streamwise); and  
\textcolor{cyan}{O} —staggered posts (spanwise).}
\label{fig:drincre}
\end{figure}

Figure~\ref{fig:drincre}a shows the mean velocity profiles shifted using the modified drag change formulation. The excellent agreement with the smooth-wall reference case demonstrates the robustness of Equation~\ref{eq:deltau} in capturing the overall drag modification induced by the filamentous surface. Note that, in the figure, the profiles for the low and mildly inclined canopies do not extend into very small values of $y_{h_{\bot}}^+$, since the tip of the canopy is used as the reference origin in the wall-normal direction.

Figure~\ref{fig:drincre}c presents the predicted drag change $\Delta U^+$ as a function of the virtual origin for turbulence $\ell_T^+$ across all simulated canopy configurations. For comparison, the figure also includes data from \citet{abderrahaman2019modulation}, who carried out direct numerical simulations of turbulent channel flows over surfaces populated with homogeneously distributed, colocated posts, representative of the transitionally rough regime.

More recently, \citet{thakkar2018direct} performed a systematic series of DNS for rough-wall turbulent channel flows and identified two sub-regimes within the transitionally rough regime. Based on the equivalent sand grain roughness height ($k_s^+$), these were classified as the lower transitionally rough regime ($0 \leq k_s^+ \leq 13.05$) and the upper transitionally rough regime ($13.05 < k_s^+ < 78.3$). These benchmarks provide a useful basis for comparing the drag behaviour of our inclined canopies with canonical rough-wall flows.
The configuration with $\theta = 77^{\circ}$ aligns closely with the trend observed for surfaces populated with colocated and staggered posts, suggesting that this canopy behaves similarly to a transitionally rough surface. Based on the magnitude of $\Delta U^+$, we hypothesise that this configuration falls within the lower transitionally rough regime, where drag modulation is still primarily governed by viscous forces \citep{thakkar2018direct}. This interpretation is further supported by the collapse of the r.m.s. velocity fluctuation profiles onto those of the smooth open-channel flow (see Figure~\ref{fig:dreduction}), indicating that the turbulence structure remains largely unaltered and reminiscent of a smooth wall flow.

In contrast, decreasing the inclination angle results in deeper turbulence penetration and, for the same value of $\ell_T^+$, the mildly and weakly inclined canopy cases exhibit larger values of drag change than their roughness analogues. This marks a transition from a rough-wall regime to a canopy-dominated regime, where the drag is no longer purely characterised by roughness metrics.

An alternative approach, proposed by \citet{orlandi2008direct}, reveals a strong linear relationship between drag change and the wall-normal velocity fluctuations—i.e., the transpiration intensity $v_h^{\prime+}$—at the texture tip. This formulation effectively decouples the drag change from specific geometric features such as texture height, density, or shape. Figure~\ref{fig:drincre}d presents $\Delta U^+$ as a function of $v_h^{\prime+}$, showing excellent collapse across all configurations and confirming that transpiration intensity reliably captures the increasing drag associated with larger or more intrusive surface elements. A similar correlation was also reported by \citet{rosti2017numerical} in the context of turbulent flows over hyper-elastic walls.

These findings indicate that the transition from the roughness regime to a canopy-like regime is fully characterised by the enhanced wall-normal transpiration at the canopy tip, rather than by geometric parameters alone.
This behaviour is also reflected in the non-linear term—namely, the transpirational virtual origin $\ell_N^+$—of the proposed modified drag change model (see Figure~\ref{fig:drincre}b). 
Specifically, $\ell_N^+$ generally increases with increasing $\ell_T^+$, reaching a maximum for the mildly inclined configurations. Beyond this point, further increases in $\ell_T^+$ lead to a non-monotonic behaviour and eventual asymptotic saturation of $\ell_N^+$. This trend reflects the composite definition of $\ell_N^+$, which depends on the interplay between mean shear and wall-normal transpiration at the canopy tip. The saturation indicates that, despite further geometric changes, the additional drag effects captured by $\ell_N^+$ become less sensitive to inclination, signalling the onset of the fully developed canopy regime. A similar behaviour was documented in the seminal work of \citet{nikuradse1933stromungsgesetze}, who investigated pipe flows with internal surfaces roughened by sand grains of height $k_s^+$. By systematically varying $k_s^+$, he captured the full transition from hydrodynamically smooth to fully rough conditions.

Nikuradse introduced a logarithmic law for the velocity profile in rough-wall flows, commonly expressed as $U^+ = A + 5.75\log_{10}(r/k_s)$, where the influence of surface roughness is encapsulated in an additive constant $A$. Notably, this constant reaches a peak within the transitionally rough regime and saturates to a value of approximately $A \approx 8.5$ in the fully rough limit \citep{thakkar2018direct}. The evolution of $\ell_N^+$ with $\ell_T^+$ mirrors this behaviour, supporting the interpretation that the transition into the canopy regime is analogous to the progression into the fully rough regime in classic roughness studies.

We conclude by examining the total drag experienced by the flow. All simulations in this study were performed at the same nominal bulk Reynolds number, $Re_b = 6000$, which corresponds to a constant volumetric flow rate. This condition facilitates a direct comparison between the smooth-wall reference and the inclined-canopy configurations, allowing the total drag to be inferred from the mean pressure gradient.
The drag reduction is quantified by the ratio
\(
DR = \frac{G_{px} - G_{px,s}}{G_{px,s}},
\)
where $G_{px}$ and $G_{px,s}$ denote the mean streamwise pressure gradients for the inclined-canopy and smooth-wall configurations, respectively. For the $\theta = 90^\circ$ case, we observe a drag reduction of $DR = 3.2\%$, whereas for $\theta = 77^\circ$, the drag reduction is $DR = -12.1\%$ (here, the negative sign indicating an increase in drag relative to the smooth-wall case.)
The limited drag-reducing performance of the fully inclined configurations may be attributed to the non-linear breakdown of drag-reduction regimes observed in riblet-inspired surfaces when the protrusion height exceeds the optimal viscous-scaled threshold \citep{garcia2011hydrodynamic}.

\subsection{Turbulent structures}

We now analyse the turbulent structures characterising the various flow configurations considered in this study. The structures are analysed through the spectral decomposition of the fluctuating velocity components. Since we have previously established that the flow over the extremely inclined canopy ($\theta=90^{\circ}$) closely resembles that over a smooth wall, results for this configuration are omitted for brevity.

As anticipated earlier, to extract the fluctuating velocity field, we employ a triple decomposition. Following \citet{nikora2007doublebac}, we write:
\begin{equation}\label{eq:tridecom}
    u(x,y,z,t)=\langle \overline{u}(y)\rangle + \tilde{u}(x,y,z) + u'(x,y,z,t),
\end{equation}
\begin{equation}
    \tilde{u}(x,y,z)=\overline{u}(x,y,z) - \langle \overline{u}(y)\rangle,
\end{equation}
where the instantaneous velocity $u(x,y,z,t)$ is decomposed into three components: a time- and space-averaged mean $\langle \overline{u}(y)\rangle$, a time-averaged spatially varying (dispersive) component $\tilde{u}(x,y,z)$, and a fully fluctuating space-time component $u'(x,y,z,t)$. The dispersive velocity $\tilde{u}(x,y,z)$ captures the spatial inhomogeneities associated with the filament geometry and vanishes in the homogeneous outer flow, where the decomposition reduces to the classical Reynolds' formulation.

\begin{figure}
\centering
\includegraphics[width=0.8\textwidth]{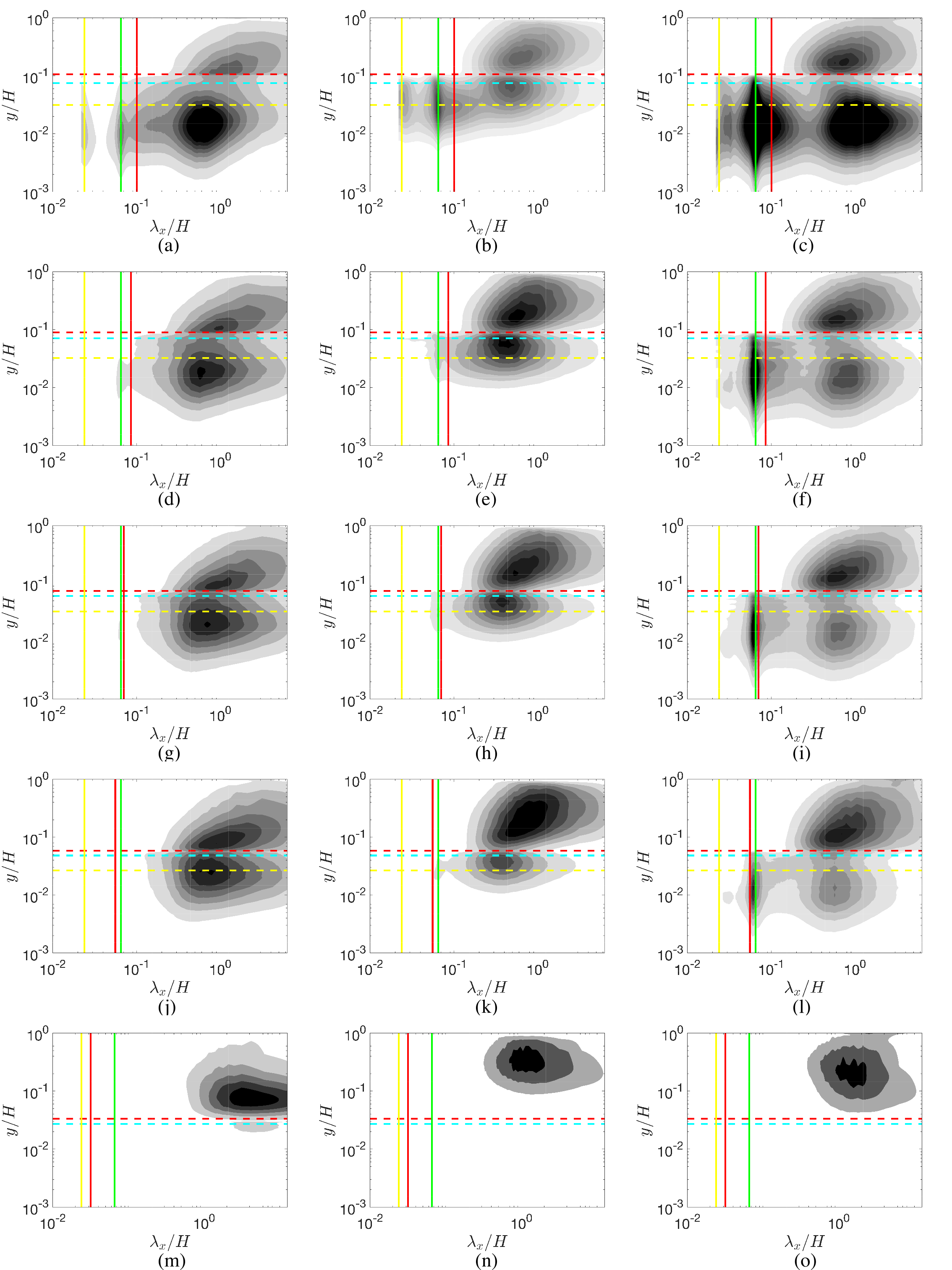}
\caption{One dimensional pre-multiplied spectra as a function of the wall normal ($y/H$) direction and the streamwise wavelength ($\lambda_x/H$). The rows from top to bottom represent increasing values of $\theta$. The columns from left to right represent $k_x\phi_{u'u'}/u_{\tau,l}^2$ with a range of [0, 0.8] and increments of 0.1, $k_x\phi_{v'v'}/u_{\tau,l}^2$ with a range of [0, 0.3] and increments of 0.02, and $k_x\phi_{w'w'}/u_{\tau,l}^2$ with a range of [0, 0.4] and increments of 0.04, respectively. The yellow, green and red vertical lines represents the wavelength at $\lambda_x/H=d$, $\lambda_x/H=\Delta S$ and $\lambda_x/H=h$. The red, yellow and cyan horizontal dashed lines represents $y/H=h$, $y/H=y_{ip}$ and $y/H=y_{vo}$.}
\label{fig:xspectra}
\end{figure}

\begin{figure}
\centering
\includegraphics[width=0.8\textwidth]{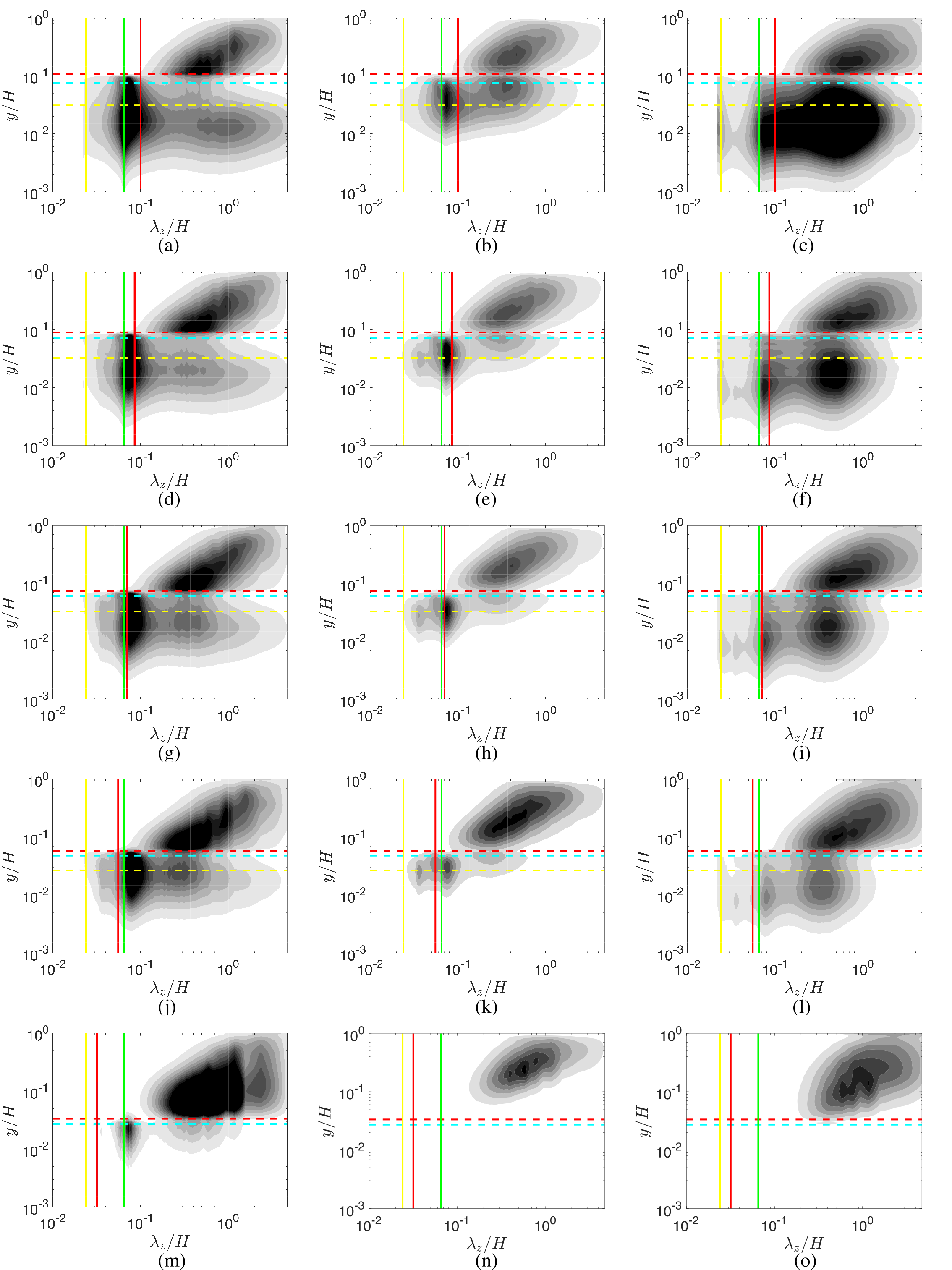}
\caption{One dimensional pre-multiplied spectra as a function of the wall normal ($y/H$) direction and the streamwise wavelength ($\lambda_z/H$). The rows from top to bottom represent increasing values of $\theta$. The columns from left to right represent $k_z\phi_{u'u'}/u_{\tau,l}^2$ with a range of [0, 1] and increments of 0.1, $k_z\phi_{v'v'}/u_{\tau,l}^2$ with a range of [0, 0.4] and increments of 0.04, and $k_z\phi_{w'w'}/u_{\tau,l}^2$ with a range of [0, 0.5] and increments of 0.05, respectively. Coloured lines have the same meaning of in Figure \ref{fig:xspectra}.}
\label{fig:zspectra}
\end{figure}

We begin our analysis by examining the spectral energy content of the velocity fluctuations in the spectral space. 
In particular, we report one-dimensional pre-multiplied energy spectra, defined as $k_i E_{\phi\phi}(k_i)$, where $E_{\phi\phi}(k_i)$ is the spectral density of the velocity component $\phi$ along the direction $i$. The pre-multiplication by wavenumber $k_i$ allows us to identify the contribution of each logarithmic interval of scales to the total energy, offering enhanced visibility of spectral peaks across a broad range of scales.
In  Figures~\ref{fig:xspectra} and~\ref{fig:zspectra} display the one-dimensional pre-multiplied spectra of the velocity components as functions of the wall-normal position and the streamwise and spanwise wavelengths, respectively. The data are organised into a $5 \times 3$ matrix of panels, where columns correspond to the streamwise, wall-normal, and spanwise velocity components, and rows represent increasing inclination angles from $\theta = 0^{\circ}$ to $\theta = 77.5^{\circ}$. Both the wall-normal coordinate and the wavelengths are non-dimensionalised by the channel height $H$, while the spectral energy densities are normalised using the local friction velocity defined in Equation~\ref{eq:locfricvelo}.

Both Figure~\ref{fig:xspectra} and Figure~\ref{fig:zspectra} show the emergence of two distinct peaks, one located within the canopy (inner layer) and one above it (outer layer). The outer peak, consistently present in all three velocity components, occurs at wavelengths $\lambda_x/H \approx 1$ and $\lambda_z/H \approx 1$, and is associated with large-scale coherent structures. In particular, the streamwise velocity fluctuations ($u'$) show a prominent outer peak just above the canopy tip, corresponding to elongated, streamwise-aligned velocity streaks. These structures are characteristic of canopy flows and have been previously documented in similar contexts \citep{monti2020genesis,nicholas2022}.

The outer peak in the pre-multiplied spectrum of $u'$ scales consistently in outer units across the different configurations, and its intensity increases monotonically with the filament inclination angle $\theta$. This trend indicates a progressive strengthening of the velocity streaks as the canopy becomes more inclined. Instantaneous wall-parallel visualisations of $u'$ in Figure~\ref{fig:dr_snap} further confirm this behaviour: as $\theta$ increases, the streaks become increasingly elongated in the streamwise direction, while their spanwise extent remains approximately constant at $\Delta z^+ \approx 100$. The longest and most coherent streaks are observed for $\theta = 77.5^{\circ}$, which was previously identified as smooth-wall-like.

\begin{figure}
\centering
\includegraphics[width=\textwidth]{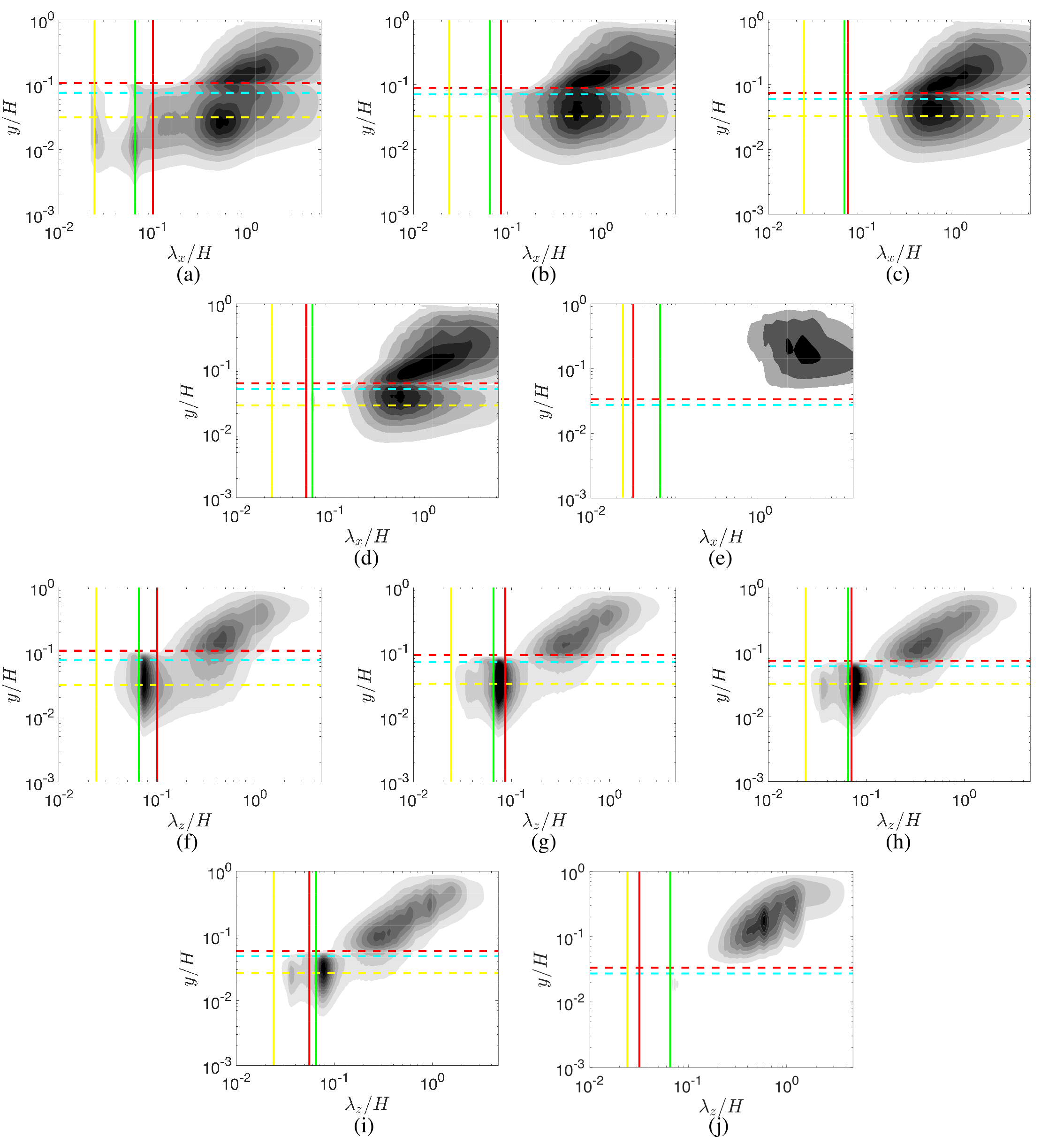}
\caption{Magnitude of one dimensional pre-multiplied cospectra ($k_x|\phi_{u'v'}|/u_{\tau,l}^2$ and $k_z|\phi_{u'v'}|/u_{\tau,l}^2$) as a function of the wall normal ($y/H$) cohordinate, the streamwise ($\lambda_x/H$) and the spanwise ($\lambda_z/H$) wavelength. The top two rows show the streamwise cospectra, the last two rows represent the spanwise ones. The panels from left to right are ordered for increasing values of $\theta$. $k_x|\phi_{u'v'}|/u_{\tau,l}^2$ has a range of [0, 0.2] and increments of 0.02, $k_z|\phi_{u'v'}|/u_{\tau,l}^2$ with has a range of [0, 0.5] and increments of 0.05; coloured lines have the same meaning of Figure \ref{fig:xspectra}.}
\label{fig:cospectra}
\end{figure}
This increase in coherence and organisation, accompanied by a reduction in meandering, resembles flow behaviour observed in drag-reducing conditions over permeable substrates \citep{rosti2018,gomez2019turbulent}.  Streaks of streamwise velocity are typically flanked by quasi-streamwise vortices, whose presence is confirmed by the outer peaks observed in the $v'$ and $w'$ pre-multiplied spectra shown in Figure~\ref{fig:zspectra}.

When analysing the cospectra in the canopy regime  for decreasing values of $\theta$, a different mechanism becomes apparent near the canopy interface, marked by the emergence of an outer peak in the $u'v'$ cospectra (Figure~\ref{fig:cospectra}). This peak is indicative of a Kelvin–Helmholtz (KH)-like instability, which is triggered by the inflection point at the canopy tip.  This instability leads to the formation of large-scale, spanwise-coherent rollers that develop near the canopy edge and generate alternating upwash and downwash of momentum into the canopy through the wall-normal velocity component ($v'$).

This interaction is clearly reflected in the pre-multiplied $v'$ spectra, which display both inner and outer peaks which are the signatures of momentum exchange across the canopy–flow interface. An additional key contributor to the development of these KH-like rollers is the non-zero transpiration at the canopy tip, as previously noted by \citet{jimenez2001turbulent}.
This is also consistent with recent findings by \citet{toedtli2024coupled}, who demonstrated that spanwise roller structures can be amplified or attenuated depending on the phase relationship between wall pressure and transpiration, highlighting the active role of wall-permeability in modulating coherent dynamics.

Although these spanwise rollers are a characteristic feature of the flow, they do not always appear as well-defined, coherent vortical structures \citep{finnigan2009turbulence}, as their signature can be masked by strong sweep events driven by large-scale motions in the outer flow \citep{monti2019large}. It should be emphasised, however, that such structures are quite ubiquitous: similar vortical features have been observed in other rough-wall configurations, including permeable substrates \citep{rosti2018} and ribbed surfaces \citep{endrikat2021influence}, where they have been linked to drag-increasing mechanisms, suggesting a common underlying dynamic.

\begin{figure}
\centering
\includegraphics[width=0.48\textwidth]{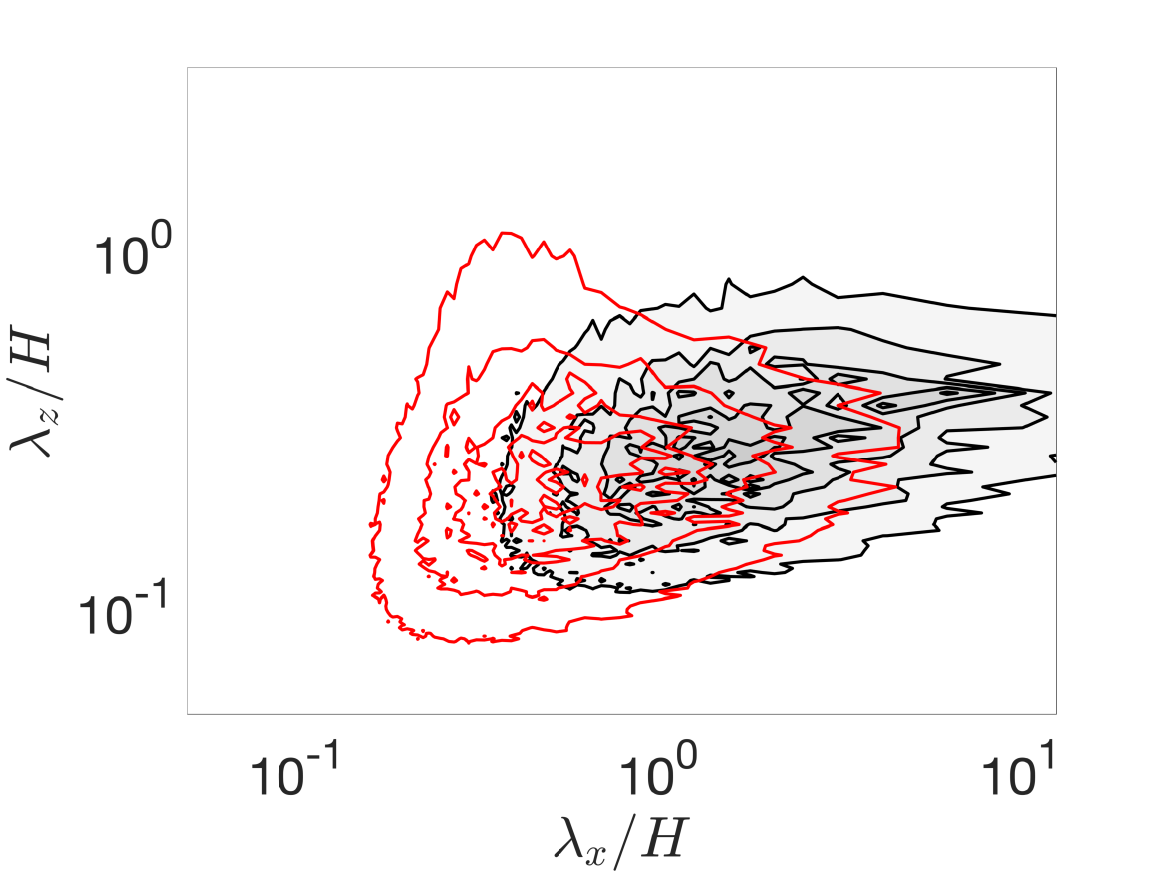}
\caption{Two dimensional pre-multiplied spectra of $k_xk_z\phi_{v'v'}/u_{\tau,l}^2$ as a function of the streamwise and spanwise wavelengths, extracted at a wall parallel location $y_{out}^+=20$. Black lines with filled contour represent the case with $\theta=77^{\circ}$, while red lines represent that with $\theta=60^{\circ}$}
\label{fig:evv}
\end{figure}

In our specific case,  as the inclination angle $\theta$ increases, the footprint of the $v'$ peak at the canopy edge progressively diminishes, eventually disappearing entirely in the $\theta = 77.5^{\circ}$ configuration. This trend suggests that the reduced transpiration caused by increasing filament inclination mitigates the KH-like instability, or, at least, suppresses the formation of spanwise-coherent rollers. This is further illustrated by the two-dimensional pre-multiplied spectrum of $v'$ in Figure~\ref{fig:evv}, which, for $\theta = 60^{\circ}$, displays a prominent peak at a large spanwise wavelength and shorter streamwise wavelength. As already mentioned, this feature is indicative of spanwise roller formation also in the context of other rough surfaces, e.g. ribbed walls \citep{garcia2011hydrodynamic}. Overall, as $\theta$ is further increased, the spectral distribution becomes dominated by long streamwise structures, reminiscent of smooth-wall behaviour. A similar transition has been observed in porous layers when reducing the wall-normal permeability \citep{rosti2018}.

\begin{figure}
\centering
\includegraphics[width=0.8\textwidth]{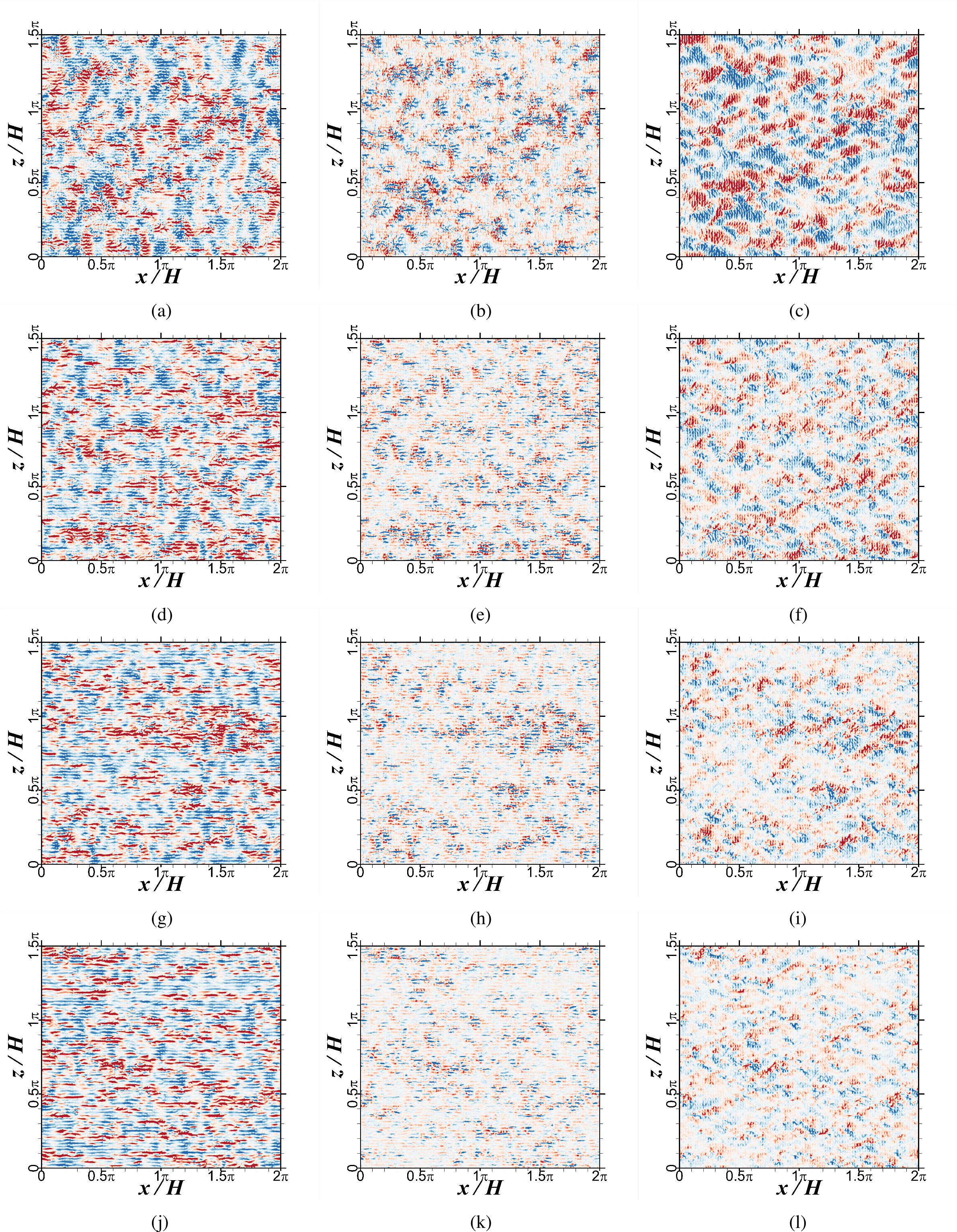}
\caption{Instantaneous realisations of the streamwise (first column), wall normal (second column) and spanwise (third column) velocity components at a wall parallel plane located at $y/H=y_{ip}$. The rows from top to bottom correspond to $\theta=0^{\circ}$, $\theta=33.5^{\circ}$, $\theta=48.25^{\circ}$ and $\theta=60^{\circ}$. 
A linear and symmetric colour bar from red to blue encompasses positive and negative values of the velocity fluctuations, made non-dimensional with the local friction velocity formulation (see Equation \ref{eq:locfricvelo}) and shown in a range $u^{\prime+}\in[-3,3]$, $v^{\prime+}\in[-2,2]$ and $w^{\prime+}\in[-3,3]$.}
\label{fig:wallpara}
\end{figure}

We next turn our attention to the internal layer of the canopy. For the characterisation of turbulence in this region, we also introduce a series of wall-parallel snapshots of the fluctuating velocity field, extracted at the internal inflection point and presented in Figure~\ref{fig:wallpara}. This figure is organised as a $4 \times 3$ matrix: rows correspond to increasing inclination angles from $\theta = 0^{\circ}$ to $\theta = 60^{\circ}$, and columns display the streamwise, wall-normal, and spanwise velocity components. While the figure spans all configurations, we begin by focusing on the upright case ($\theta = 0^{\circ}$).

In this configuration, the one-dimensional pre-multiplied spectra reveal two distinct peaks in both the streamwise and spanwise directions within the canopy. The first, located at $\lambda_x/H \approx \lambda_z/H \approx \Delta S$, corresponds to small-scale contributions from meandering motions between the filaments. Additionally, the top-row panels of Figures~\ref{fig:xspectra} and \ref{fig:zspectra} show a second small-scale contribution, likely originating from filament wakes, with a characteristic wavelength of $\lambda_x/H \approx \lambda_z/H = \mathcal{O}(d/H)$. In the fully dense regime, these contributions become more prominent due to the relatively small spacing $\Delta S/d \approx 6$.

A second, broader spectral peak is observed at $\lambda_x/H \approx \lambda_z/H \approx 1$, corresponding to large-scale structures from the outer flow penetrating into the canopy. While the length scales deep inside the canopy are constrained by geometry, the spectral energy peaks for $u'$ and $w'$ extend below the internal inflection point. In contrast, the wall-normal extent of the $v'$ peak remains limited to the virtual origin. This separation suggests that different physical mechanisms govern the behaviour of $u'$ and $w'$ compared to $v'$, beyond the effect of weak shear near the impermeable wall.

This interpretation aligns with the findings of \citet{nicholas2022}, who showed that in fully dense canopies, strong scale separation induced by small $\Delta S$ leads to a decoupling of inner and outer layers. As a result, large-scale motions from the outer flow do not reach the canopy bed but remain above the virtual origin (highlighted by cyan dashed lines in Figures~\ref{fig:xspectra} and \ref{fig:zspectra}). This is further corroborated by the velocity snapshots in Figure~\ref{fig:wallpara}, which show an absence of large-scale structures near the canopy base for the dense, upright configuration.

In contrast, sparser canopies allow large-scale structures to penetrate the full depth of the canopy, where they coexist with smaller-scale motions. However, the former tend to dominate the internal dynamics \citep{monti2020genesis}. In dense regimes, the reduced spacing $\Delta S$ acts as a geometric filter, breaking the coherence of outer structures and promoting the formation of wall-normal jets aligned with the filaments. A detailed analysis of these jets is provided in \citet{monti2022solidity}, but a brief summary is included here.

Wall-normal jets of size $\Delta S$ penetrate from the outer region and reach the lower part of the canopy, generating strong, coherent $u'$ and $w'$ fluctuations near the bed. As the flow is deflected, it creates regions of intense $u'v'$ activity associated with vortical structures that stabilise the position of the internal inflection point. This mechanism is evident in the internal peak of the $u'v'$ cospectra shown in Figure~\ref{fig:cospectra}, particularly panels (a) and (f), where the peak aligns with the mean location of the internal inflection point (marked by the yellow dashed line). The resulting momentum redistribution manifests as a spanwise modulation of $u'$ and a streamwise modulation of $w'$, shaping the structure of turbulence near the canopy bed.

We now consider how increasing filament inclination modifies this behaviour. As shown in the previous subsections, the level of penetration from the outer flow diminishes with increasing $\theta$. In addition to the geometric constraint imposed by the reduced spacing $\Delta S$, inclined filaments further obstruct wall-normal transport, limiting both the coherence and vertical extent of outer-layer structures \citep{monti2022solidity}. These effects are expected to be most pronounced in the wall-normal velocity component $v'$, which is directly tied to the degree of penetration. This is confirmed by the pre-multiplied spectra and cospectra presented in Figures~\ref{fig:xspectra}, \ref{fig:zspectra}, and \ref{fig:cospectra}.

Apart from the most extreme inclination, the $w'$ spectra remain largely consistent across configurations, showing two peaks within the canopy layer. However, clear differences emerge in the $v'$ component. The $u'$ spectra show only minor modulation, primarily in the spanwise direction. In panels (e), (h), and (k) of Figures~\ref{fig:xspectra} and \ref{fig:zspectra}, we observe a progressive attenuation of the $v'$ spectral peak at $\lambda_x/H \approx \lambda_z/H \approx \Delta S$ with increasing $\theta$. The disappearance of this feature—alongside the absence of a corresponding $u'$ peak—suggests that wall-normal jets are increasingly obstructed by the inclined canopy elements.

Rather than reaching the bed as in the upright case, these jets are intercepted by the inclined filaments and deflected along the spanwise direction. In doing so, they disrupt the bi-periodic flow around the stems and intermittently push it toward the wall. This interaction generates intra-canopy $u'w'$ structures, giving rise to the leftmost peak of the $w'$ cospectra (see third column in Figure~\ref{fig:xspectra}). Deep inside the canopy, we also observe a systematic decline in spectral energy across all velocity components (Figures~\ref{fig:xspectra} and \ref{fig:zspectra}), indicating reduced momentum transport due to the suppression of wall-normal jets. Meanwhile, the spectral peak above the canopy intensifies with increasing inclination, consistent with a growing detachment of the outer-layer dynamics.

These trends are visually supported by the wall-parallel snapshots in Figure~\ref{fig:wallpara}. The second column clearly shows the progressive weakening of $v'$ fluctuations with increasing inclination. This attenuation in wall-normal activity leads to a corresponding reduction in spanwise and streamwise modulation, ultimately altering the coherent motions that define the canopy-bed turbulence.

Finally, we consider one of the most inclined configurations at $\theta = 77^{\circ}$, which exhibits a marked departure from the behaviour observed in the other canopy cases. For this configuration, the one-dimensional pre-multiplied spectra show no distinct peaks within the canopy across most components, with the exception of the $u'$ spectra, in both the streamwise and spanwise directions. Specifically, panel (m) of Figures~\ref{fig:xspectra} and \ref{fig:zspectra} reveals an internal $u'$ peak characterised by a long streamwise wavelength ($\lambda_x/H \approx 3$) and a short spanwise wavelength approximately equal to the local filament spacing ($\lambda_z/H \approx \Delta S$).

The occurrence of this anisotropic structure can be attributed to the modulation of the bi-periodic flow within the inner canopy, previously discussed. In the absence of strong wall-normal penetration from outer-layer structures, these internal streaks become the dominant feature. This behaviour is further understood in the context of the virtual origin framework introduced earlier. For highly inclined canopies, the virtual origin of the streamwise velocity fluctuations ($\ell_u^+$) lies significantly deeper than that of the turbulence origin ($\ell_T^+$), which governs the onset of the near-wall cycle captured in the $v'$ and $w'$ spectra.

The absence of internal peaks for $v'$ and $w'$, typically indicative of quasi-streamwise vortical structures, suggests that the near-wall cycle is effectively suppressed within the canopy. In contrast, the internal $u'$ peak likely reflects elongated streaky structures whose lower portions become trapped between the inclined filaments, unable to interact fully with the wall. As a result, the canopy no longer supports the canonical cycle of wall-normal ejection and sweep events but instead exhibits a modified internal dynamics dominated by streamwise-aligned streaks that remain disconnected from the outer flow.
\section{Conclusions}
\label{conclusion}

In this study, we have carried out a series of large-eddy simulations of turbulent flows in an open channel bounded by rigid cylindrical filaments fixed perpendicularly to an impermeable bottom wall, each consisting of a vertical sheath and an upper inclined section.

With a fixed bulk Reynolds number $Re_b=6000$, the flow is principally governed by the geometric properties of the canopy: the filament height $h$, the average spacing between filaments $\Delta S$, and the inclination angle $\theta$ of the upper portion relative to the vertical.

By systematically varying the inclination angle $\theta \in \{0^{\circ}, 30^{\circ}, 48.15^{\circ}, 60^{\circ}, 77^{\circ}, 90^{\circ}\}$, we adjusted the canopy’s frontal projected height and thus its solidity $\lambda$. 
This approach allowed us to comprehensively examine distinct flow regimes under consistent conditions for filament height $h$ and spacing $\Delta S$. From this analysis, we derived the following key results: \textit{i)} a statistical characterisation of turbulent flow within and above inclined rigid canopies; \textit{ii)} a framework to predict drag variations in the external boundary layer; \textit{iii)} a detailed description of interactions between internal and external flows and their modulation by filament inclination; and \textit{iv)} a spectral decomposition of the flow field, characterising the identified flow regimes.

This framework has general applicability to submerged canopy flows where streamlining effects are induced by flexibility or orientation of the elements. 
The virtual origin analysis and the drag model proposed here can therefore be applied to flexible canopies in quasi-steady states, such as stream-aligned aquatic vegetation, engineered microstructures, and passive flow control devices. 
Conversely, our model is less suited to unsteady or oscillatory scenarios in which the filament orientation varies dynamically with the flow  \cite{rota2024dynamics, monti2023collective}.
In these cases, additional modelling assumptions would be required to capture time-varying canopy geometry and its effect on drag.

We have also shown that the inclination angle $\theta$
significantly influences the near-wall flow behaviour. As $\theta$ increases, the canopy geometry introduces alternating regions of slip (between filaments) and no-slip (at filament surfaces) conditions along the spanwise direction, particularly near the virtual origin. Simultaneously, the frontal projected area of the canopy is progressively reduced, leading to a corresponding decline in the overall canopy drag.

An increase in $\theta$ gives rise to two distinct flow regimes, as revealed by the behaviour of the roughness function
$\Delta U^+$. The \textit{canopy regime}, observed for low to moderate inclination angles ($\theta \lesssim 77^\circ$), 
is characterised by large values of the roughness function
(i.e., $\Delta U^+ \gg0$), 
in line with classical canopy flow behaviour. As the filaments become more inclined, the drag continues to diminish, eventually transitioning into the \textit{roughness regime}, associated with highly inclined canopies ($\theta \gtrsim 77^\circ$). In this regime, the canopy acts more like a transitionally rough surface, where the roughness function attains lower—but still non-zero—values, reflecting the reduced influence of the canopy elements on the external flow.\\
In particular, as the inclination angle approaches its maximum ($\theta=90^{\circ}$), the canopy behaviour departs markedly from the canonical regime. In this limit, the filaments act increasingly like roughness elements, with the inclination angle effectively controlling the displacement of the virtual origin beneath the canopy tips. These configurations resemble transitionally rough surfaces \citep{thakkar2018direct}, where both viscous and pressure drag contribute to the total resistance. Notably, a moderate drag reduction is observed at $\theta=90^{\circ}$, indicating that the canopy-induced drag becomes comparable to the skin-friction drag of a smooth wall. This suggests a transitional condition between the two most inclined cases ($\theta=77.5^{\circ}, 90^{\circ}$), with the drag increase/decrease shift occurring at an estimated angle of $\theta \sim 84^{\circ}$.

Motivated by this behaviour, we interpret highly inclined canopies as textured surfaces that displace near-wall turbulence upwards. Building on the framework proposed by \citet{ibrahim2021smooth}, we introduce a virtual origin model to predict drag changes by identifying an effective origin for turbulence. When the wall-normal coordinate is aligned to this turbulence origin, the mean velocity profile exhibits smooth-wall-like behaviour in the highly inclined limit. However, we find that using only the displacement between the mean flow and turbulence origins leads to an underestimation of the actual drag. 

We propose an additional nonlinear correction term derived from virtual origins 
of all velocity components. While this correction accurately captures $\Delta U^+$ 
even within the canopy regime, we note that the transpiration velocity $v_h'^+$ 
provides a more direct and physically transparent measure of the underlying 
transpiration physics.

In this context, the effective solidity $\Lambda_{\mathrm{eff}}$, derived from the transpiration and streamwise velocity statistics, emerges as a robust parameter to quantify the degree of coupling between the internal and external layers. 
The proposed framework allows for a unified interpretation of drag transitions and coherent structure evolution across different canopy configurations. 
Moreover, the virtual origin parameters introduced in this work serve as a bridge linking the flow structure to the canopy-induced drag, offering a useful tool to model canopy flows where filament inclination varies due to passive or active mechanisms.

The physical distinctions between regimes are further reinforced by spectral analysis of velocity fluctuations, which reveals marked differences in the coherent structures populating the flow. At low to moderate inclination angles, the outer region remains populated by spanwise-coherent motions and logarithmic structures, driven by a Kelvin–Helmholtz-like instability arising from the sharp drag discontinuity at the canopy tips. In contrast, these spanwise-coherent features vanish for highly inclined filaments, and the flow increasingly resembles that over a smooth wall, with turbulence structures shifted upwards.

As noted by \citet{nicholas2022}, in this highly inclined limit the inner and outer layers become almost fully decoupled, communicating primarily through intermittent wall-normal jets of size comparable to the filament spacing $\Delta S$. These jets emerge from the breakdown of large-scale coherence and interact with the impermeable wall to generate correlated fluctuations in the streamwise ($u'$) and wall-normal ($w'$)
velocity components. However, for the most inclined cases, this mechanism is geometrically constrained and thus significantly weakened. The filaments effectively shelter the region near the wall, reducing canopy transpiration and limiting momentum redistribution into the lower canopy.

\section{Declaration of interests}
The authors report no conflict of interest.

\section{Author ORCIDs}
\noindent
Shane Nicholas \orcidA{}: \url{https://orcid.org/0000-0002-6978-8313} \\
Alessandro Monti \orcidB{}: \url{https://orcid.org/0000-0003-2231-2796} \\
Giulio Foggi Rota \orcidC{}: \url{https://orcid.org/0000-0002-4361-6521} \\
Marco Edoardo Rosti \orcidD{}: \url{https://orcid.org/0000-0002-9004-2292} \\
Mohammad Omidyeganeh \orcidE{}: \url{https://orcid.org/0000-0002-4140-2810} \\
Alfredo Pinelli \orcidF{}: \url{https://orcid.org/0000-0001-5564-9032}

\begin{acknowledgments}
The simulations were performed using ARCHER2, the U.K. National Supercomputing Service (http://www.archer.ac.uk) through the U.K. Turbulence Consortium grant EPSRC EP/L000261/1. 
M.E.R. is supported by the Okinawa Institute of Science and Technology Graduate University (OIST) with subsidy funding from the Cabinet Office, Government of Japan. 
\end{acknowledgments}

\bibliography{pof}

\end{document}